\documentclass{aa}  
\usepackage{graphicx}
\usepackage{txfonts}
\usepackage{amsmath}
\usepackage[english]{babel}
\usepackage{gensymb}

\newcommand{\tbf}{}

\title{\texttt{PAStar}: a model for stellar surface from the Sun to active stars}

\begin{document} 
   \author{Antonino Petralia\inst{1}, 
   Jes\'us  Maldonado\inst{1},
   Giuseppina Micela\inst{1}
    }
   \institute{INAF - Osservatorio Astronomico di Palermo, Piazza del Parlamento 1, 90134 Palermo, Italy   \\
        \email{antonino.petralia@inaf.it}     }

   \date{}

\abstract 
 {The characterization of exoplanets requires a good description of the host star. Stellar activity acts as a source of noise which can alter planet radii as derived from the transit depth or atmospheric characterization.}
   {Here, we propose \texttt{PAStar}, a model to describe photospheric activity in the form of spots and faculae which could be applied to a wide range of stellar observations, from photometric to spectroscopic time series, to be able to correctly extract planetary and stellar properties.}
 {The adopted stellar atmosphere is a combination of three components, the quiet photosphere, spots and faculae. The model takes into account the effects of star inclination, doppler shifts due to stellar rotation as well as for limb darkening, independent for each component. Several synthetic products have been presented to show the capabilities of the model.}
   {The model is able to retrieve \tbf{the input surface inhomogeneities configuration} through photometric or spectroscopic observations. The model has been validated against optical solar data and compared to \tbf{alternative stellar surface activity models; e.g. \texttt{SOAP} code}. The Sun is a unique laboratory to test stellar models because of the possibility to relate unambiguously flux variations to surface inhomogeneities configuration. This validation has been done by analyzing a photometric time series from the Variability of Solar Irradiance and Gravity Oscillations (VIRGO) photometer on board of Solar and Heliospheric Observatory (SOHO) mission. Results have been compared to real solar images from the Helioseismic and Magnetic Imager (HMI) on board of Solar Dynamics Observatory (SDO) to confirm the goodness of the results in terms of surface inhomogeneities position and dimensions.}
   {The description of stellar activity is a fundamental step in several astrophysical contexts and it is covered by the method we have presented. Our model offers a flexible and valuable tool to describe the activity of stars when it is dominated by spots and faculae.}


   \maketitle

\section{Introduction}

Since the first exoplanets discoveries \citep{1992Natur.355..145W,1995Natur.378..355M}, thousands of exoplanets have been found and the focus of the scientific community has moved from their discovery to their characterisation.
Common methods as transits \citep[e.g.][]{Deeg2018} and radial velocities \citep[e.g.][]{Wright2018} to derive planets properties require the analysis of the stellar signal to correctly detect, and thereafter to characterize, \tbf{exoplanets. S}tellar activity is the main factor limiting our ability to characterize \tbf{exoplanets} or even to detect them \tbf{\citep[e.g.][]{Cameron2020}}.

The activity of a star is defined by the evolution of its magnetic field. An enhancement of the intensity or a change in the local magnetic field configuration on the stellar surface could origin different processes which alter the stellar signal in different spectral bands and/or could impact directly the properties of a nearby exoplanets throughout the so-called Star planet Interaction (SPI) processes \citep{vidotto_2019}.
One of the manifestation of stellar activity is the appearance of photospheric inhomogeneities, primarily  spots and faculae, which enhance the variability of the stellar signal being, respectively, colder and hotter of the stellar photosphere \citep[e.g.][]{Pagano2013}

In the case of transit method, one of the issues is the contamination of the stellar activity in the transit depth determination \tbf{\citep[e.g.][]{2012A&A...539A.140B,rzad009}}. When a transiting planet covers a spot or a series of spots, the difference in flux between photosphere and spot alters artificially the transit depth and shape causing an incorrect estimation of the planet radius and/or transit timing \citep[e.g.][]{2013A&A...556A..19O}. This also apply when spots effect is omitted from the out-of-transit correction of in-transit studies \tbf{\citep[][]{2009A&A...505.1277C}}.
Analogously, in transmission spectroscopy  studies, the stellar contamination affects dramatically our ability to detect chemical species in the upper atmosphere of the transiting planets \citep[e.g.][]{Edwards_2021,2024ApJ...960..107T} with spots that could mimic planetary spectral features \citep{2015ExA....40..723M,2021MNRAS.507.6118C,2021MNRAS.501.1733C,rzad009}.

The description of stellar activity is crucial to determine exoplanet parameters or even to detect them. 
Several authors have proposed different approaches to describe stellar activity. From the one hand, analytical and/or numerical models describe stellar variability in terms of spot and faculae filling factor variations without the need of knowing the position of the inhomogeneities on the stellar disk \citep[eg.][]{2012MNRAS.427.2487K,2003A&A...403.1135L}. They are extremely fast in describing the photometric variability but, due to the lack of the geometrical information, they are not able to describe spectroscopic effects. On the other hand, several models do include stellar surface reconstruction and they are able to give a deeper insight of the stellar variability at the cost of more time consuming retrieval processes \citep[eg.][]{2012A&A...545A.109B,Dumusque_2014,2016A&A...586A.131H,2020ApJ...902...73I,2024A&A...685A.173C}.
Among those, they distinguish for different geometrical description or for the inclusion of particular physical effects, eg. the blue-shift inhibition in active regions \citep{Dumusque_2014, 2016A&A...586A.131H} due to the strong local magnetic field \citep[eg.][]{1981A&A....96..345D} and spot temporal evolution \citep{2016A&A...586A.131H,2020ApJ...902...73I}. 

Here, we propose \texttt{PAStar} an alternative model to describe photospheric activity in the form of spots and faculae which represents a compromise between physical description and speed of evaluation, and, due to its flexibility, it could be easily improved. 

\tbf{It differentiates from present models, particularly, on the description of the inhomogeneities and on the geometrical construction. As for the former, the model could describe simultaneously spot surrounded by faculae, isolated facolae and isolated spot without any relation between their radii. This flexibility allows the description of complex stellar surfaces thus expanding our ability to describe and correct for the effect of the stellar activity on observed data. On the latter, the inhomogeneities and, more in general, the stellar surface are built on the projected stellar disk, therefore, avoiding approximations on the projection of the surface area elements and, consequently, improving the precision of the synthetic products.} \tbf{Moreover, the modularity of the model allows to modify and, consequently, to study in detail the effect of many aspects of the model, e.g. the limb darkening.}

The model could be applied to a wide range of stellar observations, from photometric to spectroscopic time series. The occulted flux by a transiting planet is saved to allow further and specific studies of the transmission spectrum. 

We describe the general properties and assumption of the model in Section \ref{sec:model}. We present several synthetic products in Section \ref{sec:synth_obs}, to show the capabilities of the model. \tbf{In Section \ref{sec:soap_comp} we present a comparison of photometric light curves from our model with respect \texttt{SOAP} code \citep{Dumusque_2014}, to check the consistency of the results.} In Section \ref{sec:valid} we show the validation of the model against photometric solar data while in Section \ref{sec:active star} we present a validation of the model in the case of a more active faculae-dominated synthetic star generated by \texttt{SOAP} code. Finally, in Section \ref{sec:disc} we present the discussion \tbf{and conclusions}.

\section{Model}\label{sec:model}

In our model, the stellar atmosphere is a combination of three components, the quiet photosphere, spots and faculae. Each component is characterized by considering a stellar atmosphere at a given temperature. Our framework is not bounded to any specific collection of stellar photospheric models. In this work, we \tbf{used Phoenix models, known for their ability to reproduce} the atmosphere of low-mass stars \citep[e.g.][PhD Thesis]{j_maldonado_2020_3738204}. Nevertheless, we note that other families of models, or even the assumption of a black body spectra can be considered. 

The number of the inhomogeneities is arbitrary as well as for their distribution across the stellar surface, i.e. we do not impose any preferential location. The model takes into account the effects of star inclination, doppler shifts due to stellar rotation as well as for limb darkening. Also in this case, the model is not linked to a specific limb darkening model but it require as input the limb darkening coefficient(s) as a function of the wavelength of the input stellar spectrum. Two formalisms have been implemented, linear and four coefficients formula from \cite{2000A&A...363.1081C}. 

The rotational period of star, as well as for its radius and temperature, are imposed. The model does not describe the intrinsic evolution of spots and/or faculae but only the variation in the position due to the stellar rotation.

\subsection{Geometry}

The model builds the stellar photosphere in spherical coordinates, with unitary radius and equally spaced in colatitude and longitude.

From the spherical grid (R=1, $\Theta$, $\Phi$) we compute the related Cartesian coordinate grid (X, Y, Z) as follows
\begin{equation}
 X = cos(i_{\star}-\pi/2)\cdot sin(\Theta) \cdot cos(\Phi) - sin(i_{\star}-\pi/2) \cdot cos(\Theta)
\end{equation}
 \begin{equation}
 Y = sin(\Theta) \cdot sin(\Phi)\\
\end{equation}
 \begin{equation}
 Z = sin(i_{\star}-\pi/2) \cdot sin(\Theta) \cdot cos(\Phi) + cos(i_{\star}-\pi/2) \cdot cos(\Theta)
\end{equation}

where i$_{\star}$ is the stellar inclination with respect the line of sight, $\Theta$ is the colatitude and $\Phi$ is the longitude. Star inclination ranges from 0 to $90\degree$ which correspond, respectively, to Pole-on and Equatorial view.

The inhomogeneities are assumed to be circular with the faculae not necessarily correlated to the spot, which is the case for isolated faculae. Any distortion effect on the visible disk arises from their projection on the disk, therefore, the inhomogeneities become spherical caps. This allows us to better describe their effect when they approach the limb during the stellar rotation in time series. faculae are built as circles surrounding spots which result in isolated faculae when spot radius is zero.

With these assumptions, for a given spot/facula configuration the stellar photosphere configuration is build once for all the times of a series. A window rotating oppositely to the star selects the visible surface portion (i.e. X>0) whose longitude disk center can be evaluated as follows
\begin{equation}
\phi_c(t) = -2\pi \cdot (t-t_0)/P_{\star}  
\end{equation}
where $\phi_c(t)$ is the longitude of the center of the visible disk in the stellar surface, $t$ is the time, $t_0$ is a reference time and $P_{\star}$ is the star rotational period. Therefore, the visible portion of the surface is selected considering the star inclination.

The flux from each surface element is evaluated as follows in the case of linear limb darkening
\begin{equation} \label{eq:fluxmap_lin}
\begin{split}
 F(\lambda,t,j,k)  =& Mask(t,j,k) \land \frac{f(\lambda) \cdot dA_{j,k}}{2\pi \cdot (1/2-c_1(\lambda)/6)} \cdot \\
 &\left[  1-c_1(\lambda) \cdot (1-\mu(j,k)) \right]
\end{split}\end{equation}
or in the four coefficient formalism
\begin{equation} \label{eq:fluxmap_clar}
\begin{split}
 F(\lambda,t,j,k)  =& Mask(t,j,k) \land\frac{f(\lambda) \cdot dA_{j,k}}{2\pi \cdot (1/2-\sum_{i=1}^4a_i \cdot c_i(\lambda))} \cdot  \\
 &\left[  1-\sum_{i=1}^4c_i(\lambda) \cdot (1-\mu(j,k)^{i/2}) \right]    
\end{split}\end{equation}
where $j$ and $k$ are, respectively, the indexes of colatitude and longitude, $t$ is the time, $Mask(t,j,k)$ is a matrix of integers to select accordingly the proper flux ($f(\lambda$)) based on the element area occupation. $\mu(j,k)$ is the cosine of the heliocentric angle computed as $\sqrt{1-Y^{\rm 2}(j,k) -Z^{\rm 2}(j,k)}$.
Coefficients $a_i$ are normalisation constants of the four coefficient limb darkening formula whose values are $a_1=1/10$, $a_2=1/6$, $a_3=3/14$, $a_4=1/4$\tbf{, obtained by integrating the limb darkening profile over the stellar disk to give the total disk integrated flux \citep[eg.][]{2021MNRAS.507.6118C}.}
Coefficients $c_i(\lambda)$ are the limb darkening coefficients and $dA_{j,k}$  is the projected surface elements area, which is computed as follows
\begin{equation}
dA_{j,k} = 1/2 \cdot |dl1 \cdot (dl2-dl3)-dl4 \cdot (dl5-dl6)|
\end{equation}
\begin{equation*}
\begin{split}
&dl1 = Z_{j,k+1}-Z_{j,k} \\
&dl2 = Y_{j+1,k+1}-Y_{j,k+1}\\
&dl3 = Y_{j,k}-Y_{j+1,k}\\
&dl4 = Y_{j,k+1}-Y_{j,k}\\
&dl5 = Z_{j+1,k+1}-Z_{j,k+1}\\
&dl6 = Z_{j,k}-Z_{j+1,k}
\end{split}
\end{equation*}
where $Z_{j,k}$ and $Y_{j,k}$ are cartesian coordinates of the trapezoid vertices on the projected cartesian grid. 

The flux from the single surface element is then added to a specific spectral bin, depending on the doppler shift due to the stellar rotation. Therefore, to evaluate the wavelength shift the projected disk is divided in strips whose width ($\Delta Y$) is determined by the binning of the input spectrum as in the following
\begin{equation}
    \Delta Y(\lambda) = \Delta\lambda / \delta \lambda
\end{equation}
with $\delta\lambda$ given by 
\begin{equation}
\delta \lambda = sin( i_{\star}) \cdot 2\pi \cdot Y/(c \cdot P_{\star}) \cdot \lambda
\end{equation}
where $\Delta\lambda$ is the binning of the spectrum, $c$ is the speed of light and $P_{\star}$ is the rotational period of the star. Finally, the flux from each surface element is added to the corresponding spectral bin.
This model ca be used to generate forward models or to retrieve stellar inhomogeneities properties from observations.

\section{Synthetic observations}\label{sec:synth_obs}
\begin{figure*}
    \centering
    \includegraphics[scale=0.5]{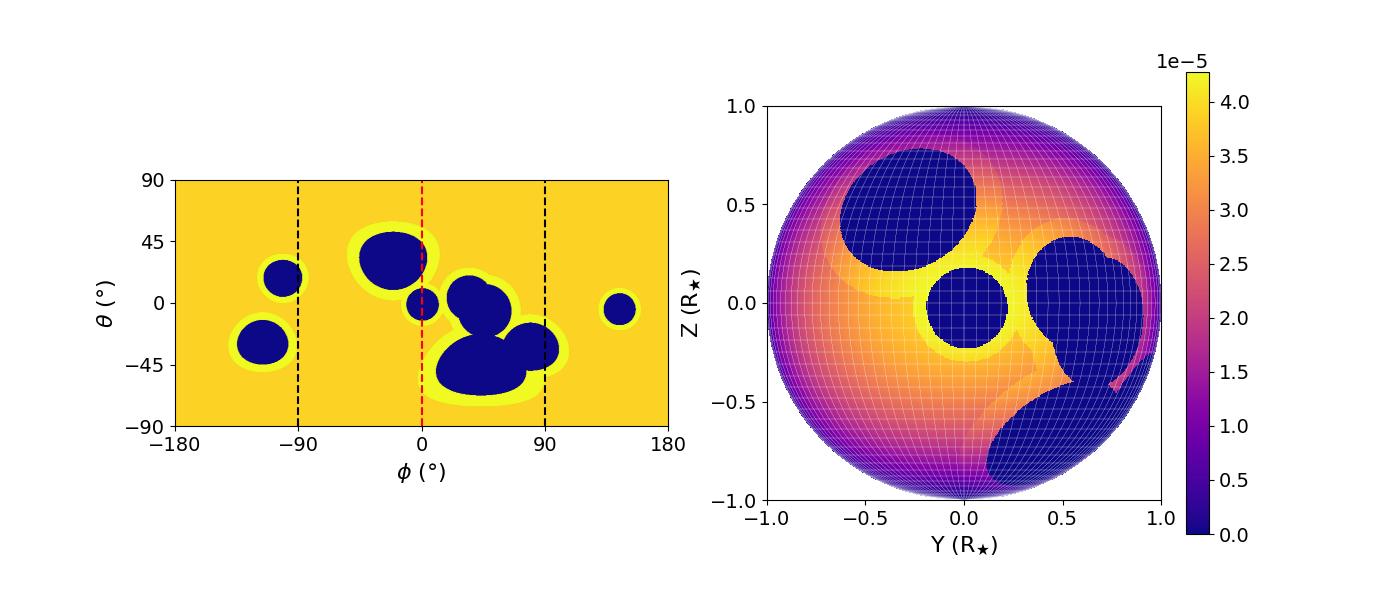}
    \caption{Configuration of inhomogeneities on the stellar surface. (Left) \tbf{Flux on the latitude/longitude \tbf{($\theta/\phi$)} domain coloured-coded ranging from 0 (in dark violet, corresponding to the presence spots), 1 (in orange, corresponding to the photosphere), to 1.1 (in bright yellow, corresponding to the presence of faculae)}. A dashed red line marks the center of the visible disk while the two adjacent dashed black lines mark the boundary of the observed disk. (Right) Projection of the left mask in the visible disk in which the colour of each surface element accounts for the \tbf{(left)} flux and the pixel area, but not for the limb darkening. White lines mark a low resolution spherical grid of 3.6$\degree$, which is 10 times lower \tbf{than} the actual grid spacing.}
    \label{fig:comb-geom}
\end{figure*}

In the following, we present several possible synthetic models generated as described in Section~\ref{sec:model}: a photometric light curve (Subsection \ref{subsec:photcurv}), a planetary transit (Subsection \ref{subsec:transit}), reproducing HD 209458 b planet characteristics \citep{2016MNRAS.463.1400D}, and a  high resolution spectrum (Subsection \ref{subsec:hiresspec}) with a resolving power of 115000.
For this exercise, we fix the stellar characteristics, which are presented in Table \ref{tab:synth-obs} and we show the differences by considering an immaculate (quiet) photosphere, with respect two different activity scenarios. The first scenario considers that stellar activity is dominated by spots whereas the second takes also into account the effect of the faculae which encircle the spots. 
In Fig. \ref{fig:comb-geom} a stellar configuration is presented, in the case of the most complex scenario. The initial observed stellar surface of the time series has been marked by two dashed black line in the surface configuration map whose view center is selected by a dashed red line. As explained in Section \ref{sec:model}, this is a moving window which selects the observed surface portion at a specific time, in this case $t=0$. 
\tbf{The geometrical grid consists of 500 points in latitude and 1000 points in longitude and it results in a pixel size of 0.36$\degree$ which represent a good compromise between computational cost and accuracy of the solution in typical photometric and spectroscopic studies.}
In the high resolution spectra evaluation, we have selected the same configuration as for the initial time of the photometric light curve. In the transiting planet case, the first mid-transit time in the series corresponds to t=1d.

\begin{table}[]
    \caption{Input parameters to compute the synthetic observables presented in Section \ref{sec:synth_obs}, including the 10 spots and corresponding faculae. \tbf{In the case of faculae, the radial increment ($\Delta$R) with respect the spot radius has been presented.}}
    \label{tab:synth-obs}
    \centering
    \begin{tabular}{c|c}
       \hline
       \multicolumn{2}{l}{Star} \\
       \hline
       P$_{\star}$ (d)  & 4 \\
       c$_i$ & 0.6 ($\forall$ T and $\lambda$)\\
       R$_{\star}$ (R$_{\odot}$)  & 1.3 \\
       T$_{phot}$ (K) & 5000\\
       T$_{spot}$ (K) & 4000\\
       T$_{fac}$ (K)  & 5500\\
       log g$_{\star} (cm\  sec^{-2})$ & 4.5\\
       $[Fe/H]$  (dex)    & 0\\
       $[\alpha/Fe]$ & 0 \\
       i$_{\star}$ ($\degree$) & 90 \\
       \hline
       \multicolumn{2}{l}{Inhomogeneities} \\
       \hline
       $\theta$ ($\degree$) & 23.37, -29.05, -45.47, -6.02, 17.55, \\
                     & -1.47, -4.92, -32.30, 30.33, 3.23 \\
       $\phi$ ($\degree$) & -103.96, -115.83, 43.63, 46.26, -101.11 \\
                     &  0.86, 144.73, 79.64, -20.56, 34.96\\
        R$_{spot}$ (R$_{\star}$) & 0.11, 0.28, 0.38, 0.33, 0.23 \\
                             & 0.20, 0.20, 0.30, 0.36, 0.28\\
        $\Delta$R$_{faculae}$ (R$_{\star}$) & 0.04, 0.09, 0.13, 0.11, 0.08 \\
                             &  0.07, 0.07, 0.10, 0.12, 0.10\\
        \hline
    \end{tabular}
    
\end{table}

\subsection{Photometry}\label{subsec:photcurv}
\begin{figure*}
    \centering
    \includegraphics[scale=0.4]{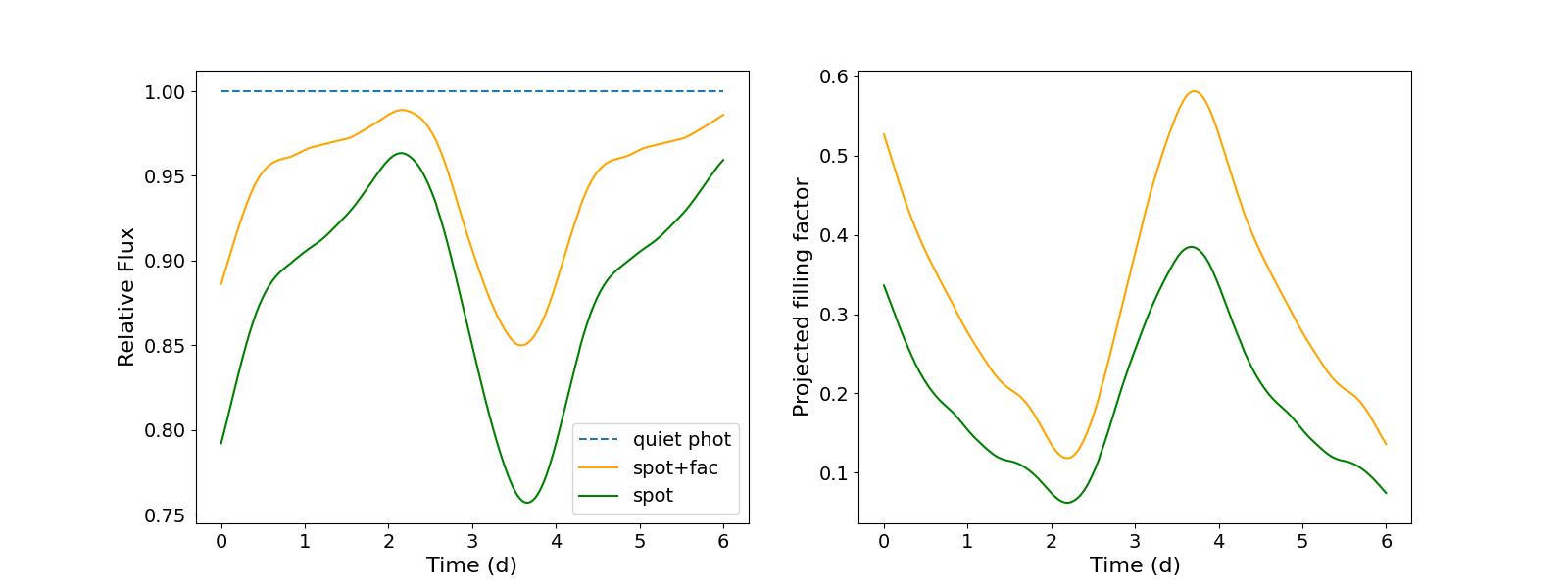}
    \caption{(Left panel) Photometric time series obtained from the model with parameters listed in Table \ref{tab:synth-obs} and considering black body emission, in the case considering both spots and faculae (orange), only spots (green) and in the case of an immaculate photosphere (dashed blue). (Right panel) Projected filling factors as a function of time for the two activity scenario considered, and color coded as on the left panel.}
    \label{fig:phot_synth}
\end{figure*}

Here, we present a synthetic photometric time series generated from the configuration shown in Fig.~\ref{fig:comb-geom}. For its evaluation, the Black Body flux $\sigma T^{4}$ has been considered for both photosphere, spots and faculae, with $\sigma$ being the Stefan Boltzmann constant and T the temperature, considering the temperatures defined in Table ~\ref{tab:synth-obs}.
In Fig.~\ref{fig:phot_synth} (left), we present 6 days of the synthetic time series, corresponding to 1.5 periods of stellar rotations. The initial star configuration match the one shown in Fig.~\ref{fig:comb-geom} (right) and the star has been rotated clockwise. Together with the white light curves, in Fig.~\ref{fig:phot_synth} (right) we also present the projected filling factor as a function of time (or rotation), defined as the total covered area by inhomogeneities on the projected visible disk divided by $\pi$, i.e. the total projected area, being $R_{\star}=1$. In the case of the complex scenario, the filling factor is due to both spots and faculae.

The initial star configuration marks a condition of very high activity in which inhomogeneities cover more than half of the visible disk. As the rotation proceeds, large spots approach the limb and the flux rises substantially due to the effect of the limb darkening, reaching its maximum after half of the rotation, where only few small spot is visible on the disk with a projected filling factor of $\sim0.1$. At later times, a conglomerate of inhomogeneities approaches the visible disk increasing drastically the filling factor (up to 0.6) while the flux alternatively falls towards its minimum at t=3.7d. At t=4d the star completes a rotational period and the configuration and, therefore, the time series repeat themselves.

The presence of the faculae affects substantially the flux. Although faculae have only built with a increment radius which is 1 over 3 the radius of the spots, their higher temperature with respect to the spots, but also with respect to the photosphere, compensates for their smaller filling factor leading to a maximum increase in the total flux of $\sim10\%$, with respect of the spot case, when their projected filling factor is $0.2$. 

\subsection{Photometric planetary transit}\label{subsec:transit}

Here, we present the synthetic transit of a planet reproducing the characteristic of HD 209458 b but facing our synthetic active star. We have considered the Black Body fluxes for the stellar and inhomogeneities fluxes as it has been done for the photometric time series in Section~\ref{subsec:photcurv}. 
To evaluate the synthetic transit \texttt{PAStar} requests the coordinates of the planetary orbit in the projected disk, i.e. Y and Z coordinates, and the ratio of the planet to the stellar radius (R$_p$/R$_\star$). All of these information have been set straightforward from the python package \texttt{PyLightcurve v4.0.1}~\citep{2016ApJ...832..202T} which allows to evaluate directly the orbits (and many other properties) of all the planets in the {Exoplanet Characterisation Catalogue} within the {ExoClock} project \citep{2022ExA....53..547K}, which also contains HD 209458 b. Although our synthetic star radius (R=1.3R$_\odot$) differs slightly with respect to the one of HD209458 \citep[R=1.2R$_\odot$, e.g.][ and references therein]{2016MNRAS.463.1400D}, for the sake of simplicity, we do not change our star parameters which results in an overestimate of the planetary radius (i.e. $\sim10\%$) being the R$_p$/R$_\star$ fixed. Although this discrepancy would give uncorrect results in a specific study, here we aim at showing the general effect of spots and faculae crossing events, therefore, the exact value of planetary radius is not relevant.
In this configuration, during the transit the star rotates of $\sim$10$\degree$. For this reason, we do consider in the evaluation the evolution of the inhomogeneities due to rotation.

\begin{figure*}
    \centering
    \includegraphics[scale=0.4]{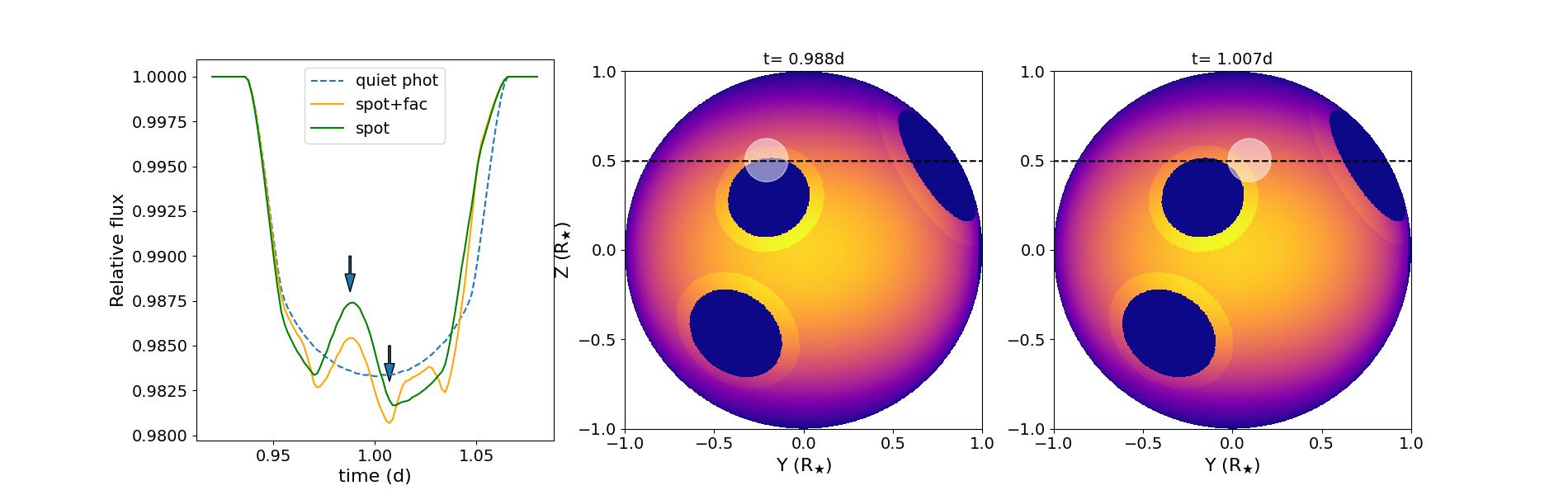}
    \caption{(Left) Transit of HD 209458 b in the case of (dashed blue) quiet photosphere, (green) spot-only activity and (orange) spot+faculae scenario. Fluxes are normalised by their out-of-transit fluxes. (Mid and Right) Snapshots of the stellar configuration at two times marked by the blue arrows in the left panel. The transit cord has been marked by a dashed black line while the planet position by a transparent and white circle, with the planet extent correctly scaled to the stellar radius. \tbf{Stellar flux has been coloured-coded as for Fig.\ref{fig:comb-geom}}.}
    \label{fig:transit}
\end{figure*}

In Fig.\ref{fig:transit} we present a single transit of HD 209458 b together with two snapshots of the stellar configuration at two times during the transit. As in the previous sections, we have compared the immaculate case together with the two activity cases, i.e. spots and spots+faculae. During the transit, the planet crosses two  distinct spots (and faculae) allowing us to disentangle their effects in the transit depth, the first spot is located close to the center while the second crosses the star limb. 

When the planet crosses a spot, we observe a rise of the flux during the transit which is more pronounced when faculae are not being considered. This is because spots have a lower flux with respect to the photosphere therefore their occultation rises the relative flux with respect to the average out-of-transit flux.
faculae act alternatively having a higher flux with respect of the photosphere, therefore their occultation lead to a fall of the flux during the transit. These effects in combination to a high activity cases or high spot coverage along the transit chord could lead to a wrong estimate of the transit depth and, consequently, of the planet radius. If the spot occulted is close to the limb, the high spot contrast (or low flux) leads to a distortion of the transit shape for the same effect described above, which lowers artificially the transit duration. These are quite typical observed effects \citep[e.g.][]{Bruno_2018}.

The presence of spots affects the transit depth also when they are not crossed. They lower substantially the flux during the transit with respect the value when an immaculate photosphere is being considered (see the green line in Fig.\ref{fig:transit} between the crossed spots). This effect can be mitigated when an accurate out-of-transit correction is implemented.

\subsection{High resolution spectra}\label{subsec:hiresspec}

 To calculate high resolution spectra the code makes use of predefined spectral energy distributions (SEDs) to be combined depending on the configuration of surface inhomogeneities. Here, we use Phoenix spectra from the database \cite{2013A&A...553A...6H}\footnote{\url{https://phoenix.astro.physik.uni-goettingen.de}}, with the stellar characteristics listed in Table \ref{tab:synth-obs}.
 
This database covers the wavelength range from 500 $\AA$ to 55000 $\AA$ with resolutions of 0.1 $\AA$ in the ultraviolet (UV) bands (500-3000 $\AA$), R500000 in the optical and near infrared (NIR) bands (3000-25000 $\AA$) and R100000 in the infrared band (25000-55000 $\AA$). The parameter space covers from 2300 K to 8000 K for the temperature, 0.0 to 6.0 for the logarithm of gravity, -4.0 to 1.0 for the Fe to H ratio and -0.2 to 1.2 for the $\alpha$ to H ratio. 

The high resolution of the Phoenix spectra in the optical and infrared bands results in a very computational expensive evaluation of synthetic products. For this reason, we lower the resolution of the spectra to match the one of real case studies. We chose to reproduce data from the high-resolution spectrograph HARPS-N \citep{2012SPIE.8446E..1VC} which has a resolution of R115000 in the optical band between 3800 and 6900 $\AA$. Before to perform this operation, spectra from the database has been modified in order to describe observed spectra from Earth rather than vacuum, therefore wavelengths have been shifted following \cite{1996ApOpt..35.1566C} by applying  

\begin{equation}
\lambda_{air} = \lambda_{phoenix}/f
\end{equation}

where $\lambda_{air}$ is the wavelength as if the star has been observed from ground, $\lambda_{phoenix}$ is the vacuum wavelength of the synthetic Phoenix spectra and $f$ is the correction factor \tbf{\citep{1996ApOpt..35.1566C}} given by

\begin{equation}\begin{split}
f =& 1.0 + 0.05792105/(238.0185-(10^4/\lambda_{phoenix})^2)+\\
  &+0.00167917/(57.362-(10^{4}/\lambda_{phoenix})^2)
\end{split}\end{equation}

After the correction on the wavelengths, we perform the spectrum degradation by applying a gaussian filter on the spectra. 
We use the procedure \texttt{gaussian\_filter} in the public available python package \texttt{scipy} to perform this task. Since the procedure works with pixels in images, we have to convert the sigma, needed by the filter, from angstrom to pixel. We select a very strong line, i.e. $H_{\alpha}$ absorption line ($6562.8$ $\AA$), to calculate the full width at half maximum (FWHM) and the resulting sigma by the following

\begin{equation}\begin{split}
&FWHM    = \sqrt{ (\lambda_{0}/R_{harpsn})^2-(\lambda_{0}/R_{phoenix})^2}\\
&\sigma  = FWHM/(2\sqrt{2}\ln{2})/dA
\end{split}\end{equation}
where $\lambda_{0}$ is the reference wavelength of $H_{\alpha}$ line, $R_{harpsn}$ and $R_{Phoenix}$ are the resolving power of, respectively, the HARPSN instrument and the Phoenix spectra, $dA$ is bin width of the Phoenix spectra. We have to note that this procedure makes sense only in the case of non-saturated absorption lines, therefore, in cases in which $H_{\alpha}$ absorption line is saturated an other choice should be made.

\begin{figure*}
    \centering
    \includegraphics[scale=0.45]{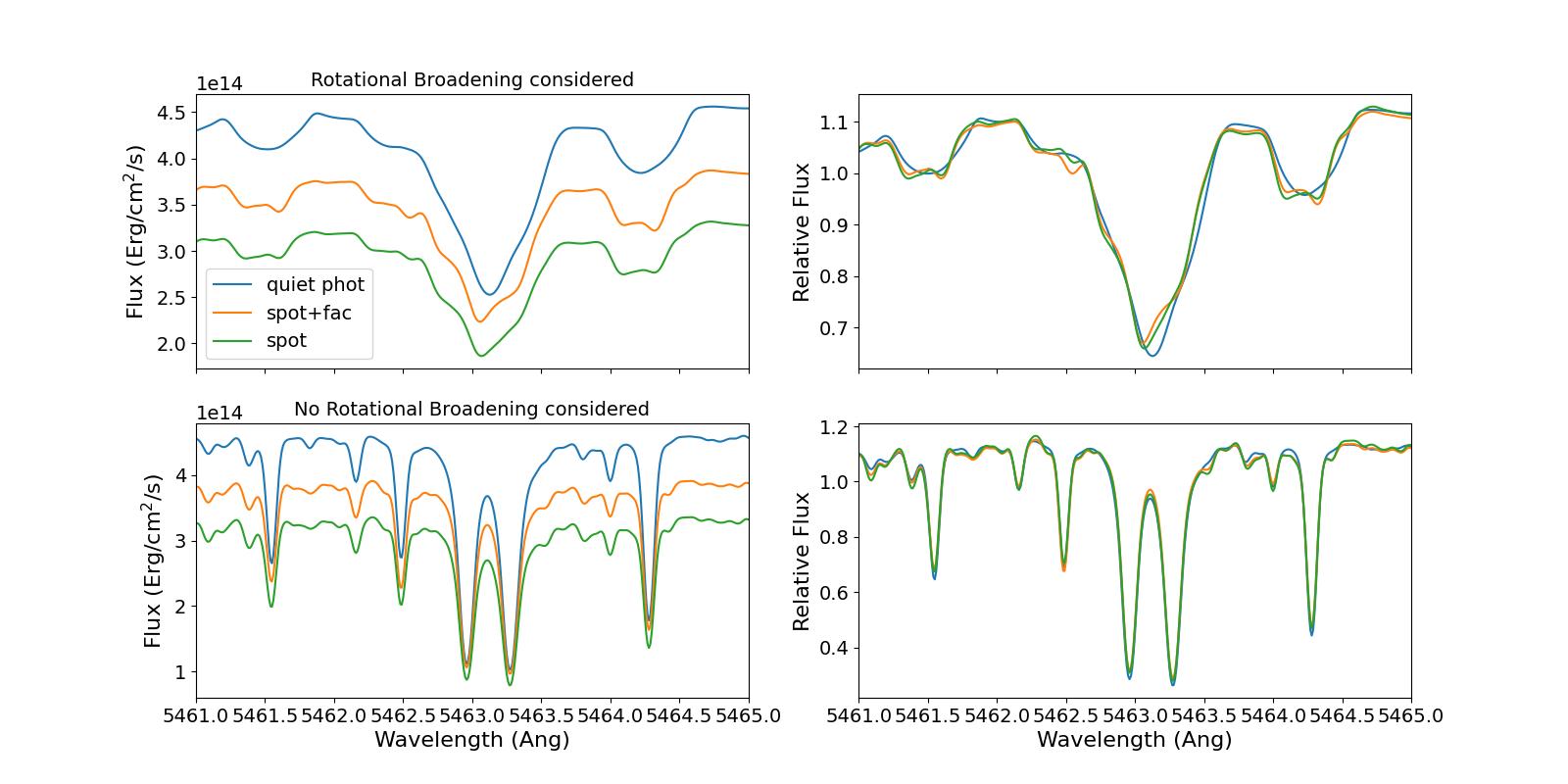}
    \caption{High resolution spectra computed by the model. (Top rows) broadened spectra by star rotation in the case of immaculate photosphere (blue), complex activity scenario (orange) and only spots scenario (green). (Bottom row) Same as for top rows but without considering the rotational broadening in the computation. For comparison purposes, absolute fluxes (left panels) are presented together to the normalized ones (right). The normalisation trend has been evaluated with a gaussian filter of 1200 $\AA$ as sigma.}
    \label{fig:hrs_range}
\end{figure*}

After the previous steps, the stellar spectrum has been calculated in the case of the configuration shown in fig \ref{fig:comb-geom} and in the two scenarios presented in the previous section, which are the complex activity (spots and faculae) case and the only spots case.
In Fig. \ref{fig:hrs_range} we present a selected and narrow spectral window of the synthetic spectrum, ranging from 5461 $\AA$ to 5465 $\AA$. This spectral windows is populated by FeI lines \citep[5461.550 \AA, 5462.959 \AA, 5463.277 \AA, 5464.280 \AA,][]{1994ApJS...94..221N}, by a NiI \citep[5462.493 \AA,][]{1993PhyS...47..628L} line and by a blending of CrI \citep[5463.928 \AA,][]{kiess1953} and NiI \citep[5463.920 \AA,][]{1993PhyS...47..628L}.

The primary effect of the inhomogeneities is to shift the level of flux in the selected range, depending on their projected filling factor. Secondarily, as can be noticed from top right panel, the line centroid has been blue-shifted by the presence of a conglomerate of spots on the right portion of the visibile disk (see Fig. \ref{fig:comb-geom}), which are coming towards the observer. When a time series is considered, this effect results in a periodic shift of the line centroid correlated to the star rotation. This is the very typical effect in which radial velocities of star are modulated by the stellar activity, in this case mostly dominated by spots.

Other important effects to be noticed are from one side, the broadening due to stellar rotation blends all absorption line which lie within the maximum doppler shift ($\sim$0.3 $\AA$ in this range or 16.5 km/s). On the other hand, the line broadening builds coherently the shift of the line centroid with respect the non rotational case (lower panels), resulting in a much more evident modulation of the centroid shift (or radial velocity modulation) in the case of high rotational speed, or lower rotational period, when a time series is being considered.

\section{Comparison with literature: \texttt{SOAP} code}\label{sec:soap_comp}

In order to check the consistency of the synthetic products of our forward model we have compared \texttt{PAStar} outputs with an alternative model present in literature: \texttt{SOAP} code \citep[v2.0,][]{Dumusque_2014}. To derive this comparison, a synthetic photometric light curve has been generated with \texttt{SOAP} code which evaluates photometric and radial velocity variations induced by active regions, in the form of spots and faculae.
\texttt{SOAP} models faculae as spatially uncorrelated bright features with respect spots with a center-to-limb temperature variation that matches the solar case \citep[eg.][]{2010A&A...512A..39M}. For this test, our synthetic star reproduces a solar-type star whose photometric variability is dominated by 4 faculae and whose characteristics have been reported in Table~\ref{tab:synth-faculae}. 

We have modify our code to employ the same center-to-limb temperature variation of faculae temperature of the \texttt{SOAP} model, as well as for the flux contrast evaluation which considers a black body radiation at 5293.4115 $\AA$ for both spots and faculae. This changes expand the flexibility of the code to adapt to different model requirements depending on the scientific cases to be studied. 

\texttt{SOAP} employs a quadratic formula for the limb darkening description whereas in \texttt{PAStar} we employ a more complex four coefficient formula \citep{2000A&A...363.1081C}. To derive a coherent comparison we match the limb darkening description by simplifying it in both model to the linear case.

\begin{figure*}
    \centering
    \includegraphics[scale=0.7]{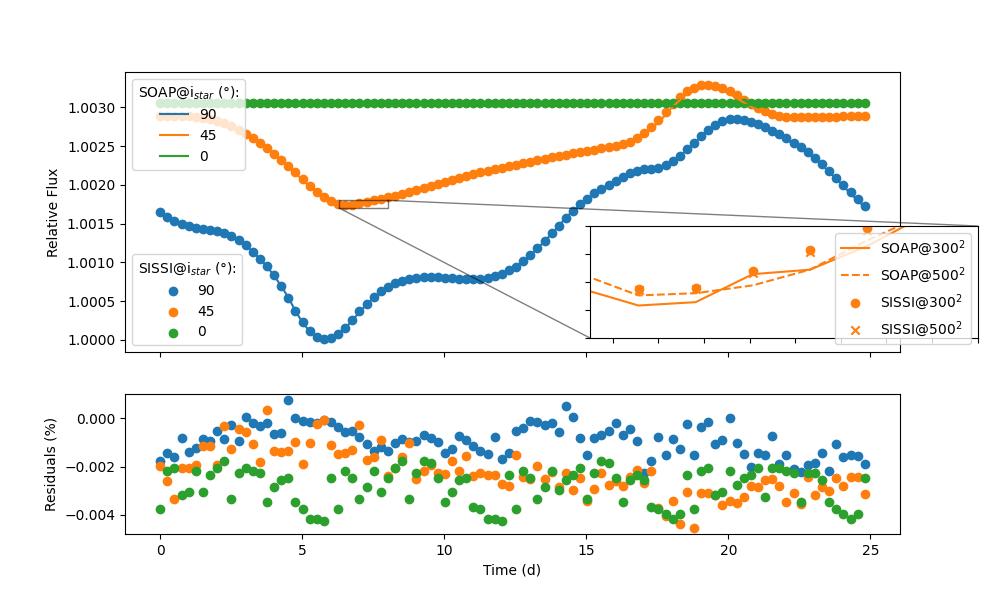}
    \caption{(Top panel) Comparison of \texttt{SOAP} forward model with respect ours with, respectively, coloured lines marking the \texttt{SOAP} solutions while circles marking ours, and using parameters listed in Table~\ref{tab:synth-faculae}. Equatorial-on solution is presented in blue, an intermediate inclination of 45$\degree$ is in orange while polar-on solution is presented in green. Fluxes have been normalised to the photospheric and unperturbed flux. \tbf{An insert in the panel show a comparison of the two models using different number of grid points to build the stellar surface. The insert follows the same notation of the main panel but two profiles are added to mark the solution obtained using 500$^2$ grid points in the case of \texttt{SOAP} code (dashed line) and ours (x marker).} (Bottom panel) Residuals of the two models evaluated as $(F_{SOAP}-F_{PAStar})/F_{SOAP} \cdot 100$\tbf{, considering the lowest grid resolution}.} 
    \label{fig:SOAP-fwmod}
\end{figure*}

Therefore, we have compared the two codes on three photometric light curves that differ on the stellar inclination, but share the same spatial configuration for the 4 faculae considered. The result of this comparison has been presented in Fig.~\ref{fig:SOAP-fwmod}. \tbf{We have also tested the difference in the modelled light curves due to the grid resolution in one of the presented cases, i.e. for $i_{\star}=45\degree$. In this case, we present the comparison using the same default number (N) of grid points that \texttt{SOAP} uses to build the visible stellar surface, i.e. N = 300$^2$, and the number we employ as a default for colatitude points, which is N~=~500$^2$. We have to remark that in our geometrical construction, when the star inclination differs from 90$\degree$, being all longitudes visible close to the pole, the resulting number of grid points in the visible disk is much higher than in the \texttt{SOAP} code due to the projection of the stellar surface on the cartesian grid.}

Differences between the two models fall below the $\sim0.004$\% threshold with our model which, systematically, gives higher fluxes. We notice a trend in the discrepancies with the amplitude of residuals which correlate with the star inclination, resulting in the polar-on solution (i$_{star}$=0$\degree$) to have the greatest deviation. This effect could arise from the different geometrical construction and/or resolution chosen in the two models, however, due to the very low amplitude of the residuals, we assert a good match between the two models. 

The difference between the two models increases at the lowest resolution. \texttt{SOAP} code benefits greatly from the increase in the grid points giving smoother profiles, but this is not requested by \texttt{PAStar} that gives nearly identical results for low and high resolution. This is due, primarily, to the geometrical construction of our code that relies on the Cartesian projected grid to integrate the stellar disk, and it results in higher precision with respect \texttt{SOAP} code at similar resolution.

\begin{table}[]
    \caption{Input parameters to compute the synthetic photometric light curve for the faculae-dominated Sun in the \texttt{SOAP} code, where $\mu$ is the cosine of the heliocentric angle (see Section~\ref{sec:model}). All other parameters has been left to their default values. Three case of stellar inclination have been considered, i$_{\star}$=0$\degree$, 45$\degree$, 90$\degree$.}
    \label{tab:synth-faculae}
    \centering
    \begin{tabular}{c|c}
       \hline
       \multicolumn{2}{l}{Star} \\
       \hline
       P$_{\star}$ (d)  & 25.05 \\
       c$_0$ & 0.6 \\
       c$_1$ & 0 \\
       R$_{\star}$ (R$_{\odot}$)  & 1 \\
       T$_{phot}$ (K) & 5778\\
       \hline
       \multicolumn{2}{l}{Faculae} \\
       \hline
       T$(\mu)$ (K)  & T$_{phot}$ + 250.9-407.7$\mu$+190.9$\mu^2$\\
       $\theta$ ($\degree$) & 30, 60, 40, 10 \\
       $\phi$ ($\degree$) & 180, 90, 40, 15 \\
       R (R$_{\star}$) & 0.15, 0.2, 0.1, 0.18 \\
        \hline
    \end{tabular} 
\end{table}

\section{Validation of the model on solar data}\label{sec:valid}

The aim of this section is to test the ability of the model to describe the stellar activity in real scientific cases in which we do not know (almost) anything about the star. However, to validate the model, we do not aim only to reproduce the star variability due to surface inhomogeneities but also their actual surface configuration (locations and sizes). Although in the stellar case there are different attempts in retrieving the spots/faculae configuration \citep[e.g.][]{1997A&A...326.1135D}, knowing the spots/faculae configuration in a unequivocal way is the chimera of the stellar science due to the lack in current instrumentation of sufficient spatial resolution. The only star in which this exercise can be done unambiguously is the Sun.  

\subsection{The Sun-as-a-Star}\label{subsec:sunasstar}
The Sun has been observed with a multitude of instruments from both space \cite[eg.][]{1991SoPh..136....1O,1995SoPh..162....1D,2007SoPh..243....3K,2012SoPh..275....3P} and ground \citep[eg.][]{1991AdSpR..11e.129S,2012SPIE.8446E..1VC}, and in many spectral bands, from Radio to X-rays. Among all the possible data available for the Sun observed as a star, we have selected a optical photometric time series from the Variability of solar IRradiance and Gravity Oscillations (VIRGO) Sun photometer (SPM) \citep{1995SoPh..162..101F,1997SoPh..170....1F,2002SoPh..209..247J} on board Solar and Heliospheric Observatory (SOHO) mission \citep{1995SoPh..162....1D}. 
Despite the simplicity of the information retrieved from a photometric time series with respect to a high resolution spectrum, it still represents a viable test to probe efficiently many aspects of the model, the geometric construction of spot/faculae, the flux evaluation and the limb darkening. Other effects such as the doppler broadening due to stellar rotation, can be probe only with a spectral synthesis, however, considering the characteristic of the Sun, such aspect can be neglected due to its slow rotation which gives a $vsini=\sim1.6$ km/s \citep{j.1365-2966.2012.20629.x}.    

VIRGO/SPM measures the spectral irradiance in different spectral ranges at the temporal cadence of 60 seconds.
We have analysed data for the measurement of the spectral irradiance at $5000$ $\AA$ with a bandwidth of $50$ $\AA$ and at the highest level of calibration (flag {L2 mission long}, publicly available at the SOHO Science Archive\footnote{\url{https://ssa.esac.esa.int/ssa/#/pages}}).     

\begin{figure*}
    \centering
    \includegraphics[scale=0.4]{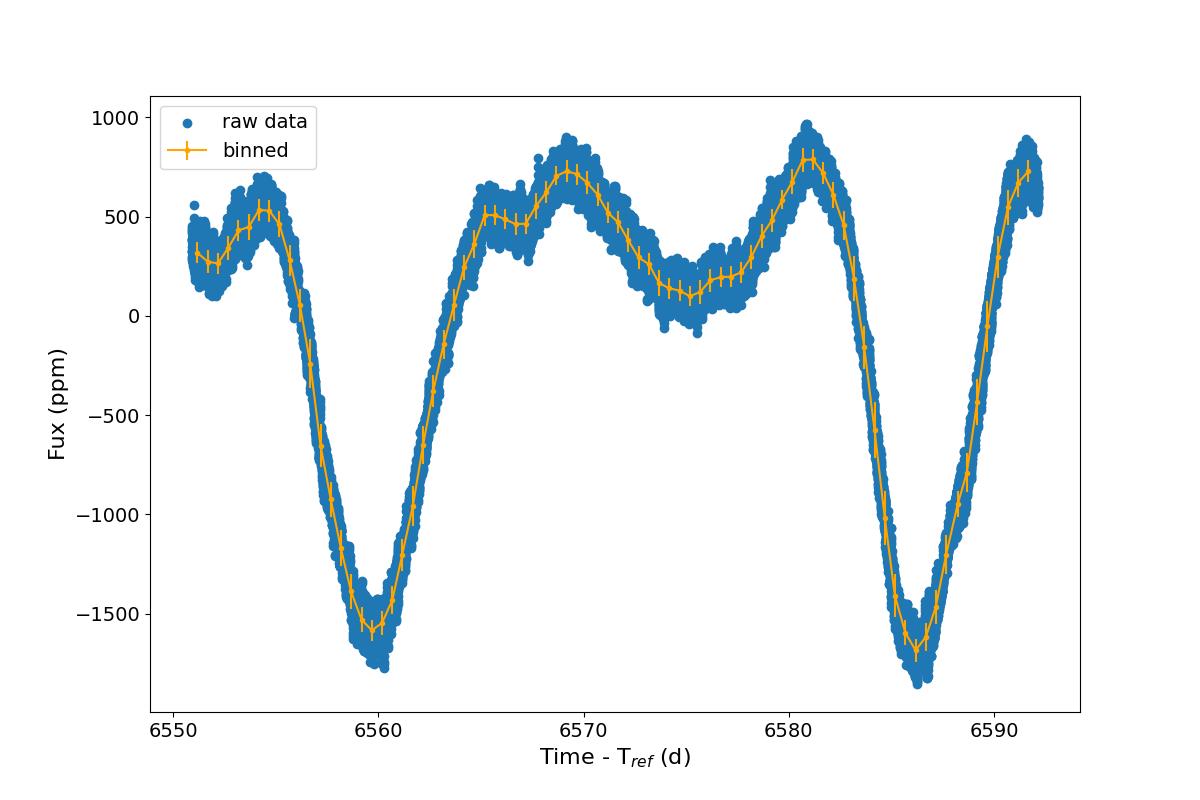}
    \caption{Solar flux in the VIRGO/SPM green channel as a function of time. Raw data are shown as blue data while binned data as dashed orange line, together with the error bars. Reference time (T$_{ref}$) is 1996-01-23T00:00:04.46 TAI.}
    \label{fig:sun_data}
\end{figure*}

Although the Sun offers unique opportunities to test and validate stellar models, it is a worst case when it comes to activity because, actually, the Sun is not an active star. Here, with the term active we refer to the filling factor of surface inhomogeneities as spots and faculae. In active stars, spots filling factor could reach, or even surpasses, the $50\%$ of the visible disk \citep[e.g.][]{O’Neal_2004} while in the Sun few percent are reached only during solar cycle maxima. \tbf{Therefore, this section doesn't represent an exhaustive validation of the model but only its first step and it is important to determine the minimum activity that our procedure can retrieve. An application of the model to active stars is presented in Section \ref{sec:active star}.  To derive this validation}, data during the 24th solar cycle maximum have been selected, in the window that starts from 2013-12-30T04:13:08.920 (International atomic time, TAI) and it extents for 41.2 days thereafter. Data in the selected temporal window are presented in Fig.\ref{fig:sun_data}. Due to the high cadence of VIRGO/SPM instrument, i.e. 60 seconds, raw data have been binned down to 12hrs. In each bin, the mean of raw data has been taken as bin value whereas the standard deviation as bin error\tbf{, corresponding to an average error of $~70$ ppm}.   

\subsection{Assumptions}\label{subsec:assumpt}
To proceed with the validation, we made the hypothesis that the Sun is dominated by dark spots in the selected temporal window, i.e. most of the variability is related by the variation in location of sunspots with zero temperature, therefore, we also neglect the effect of faculae. In the case of VIRGO green channel this choice is supported by measurements of the intensity ratio between umbra and photosphere by \cite{1984SoPh...90...17A} which show a weak variation of the ratio through the disk at 5790 $\AA$ and values below 0.1, decreasing for shorter wavelengths. Moreover, for the sake of simplicity, we neglect the inclination of the Sun with respect to the line of sight and we set i$_{\star}$ to 90$\degree$. Although these assumptions could be awkward for a well known object as it is the Sun, the spirit of the validation is to treat the Sun as a star in a typical and almost blind search for activity characteristics. Furthermore, the rotational period has been imposed to 26.5 d, retrieved from data as time between two periodic minima.

A limb darkening four coefficients formalism has been set whose values of the coefficients have been calculated with \texttt{ExoTETHyS} v2.0.10 python package \citep{2020AJ....159...75M}. The relevant parameters used for the calculation are presented in Table \ref{tab:parexot} together with the resulting coefficients. For this task, we do not set the exact values of the Sun's parameters but the closest values to a point grid of the \texttt{Phoenix\_2012\_13} \citep{2013A&A...552A..16C} database to avoid interpolation on the results.

\begin{table}[]
    \caption{
    Parameters used in the evaluation of the limb darkening coefficients. Last four lines are the resulting coefficient values.}
    \label{tab:parexot}
    \centering
    \begin{tabular}{c|c}
        \hline
         star\_effective\_temperature (K) & 5800  \\
         star\_log\_gravity & 4.5\\
         star\_metallicity & 0 \\
         passbands & uniform\_phoenix\_2012\_13\\
         limb\_darkening\_laws & claret4\\
         stellar\_models\_grid & Phoenix\_2012\_13\\
         spectral bin (\AA)& 4975-5025\\
         c1 & 0.29852834  \\
         c2 & 0.25020459\\
         c3 & 0.67279828\\
         c4 &  -0.41401905\\
         \hline
    \end{tabular}
    
\end{table}

\subsection{Retrieval framework}\label{subsec:retrf}
The model has been coupled to a retrieval framework in order to obtain parameters best-fit values and their errors. As for retrieval framework, we use \texttt{Multinest v3.10} \citep{2009MNRAS.398.1601F} to derive bayesian inference through the python interface offered by the package \texttt{PyMultiNest v2.12} \citep{2014A&A...564A.125B} and with the following likelihood

\begin{equation}
\begin{split}
        \ln{p({y_n},{t_n},{\sigma},v)} =& -1/2\sum_n \left[ (y_n-F(t_n,v))^2/\sigma^2 + \ln2 \pi \sigma^2 \right]
\end{split}
\end{equation}

where $y_n$ and $t_n$ are, respectively, data fluxes and times, $\sigma=\sqrt{\sigma_n^2+\sigma_j^2}$ with  $\sigma_n$ as data errors and $\sigma_j$ as white noise jitter term, introduced to account for unknown source of errors. Observed data are modelled as the following
\begin{equation}
    \frac{S(t)-S_0}{S_0}[ppm] = 10^{6} \cdot (f(t_n,v) \cdot f_{scale} - 1)    
\end{equation}
where S(t) is the solar flux at time $t$, $S_0$ is a flux value with respect data are shifted and scaled by the L2 Mission Long correction\footnote{from data header: Corrected from orbit, degradation, outliers, attractors; Polynomial fit (7 degrees) and 2 month highpass filter applied}, $ppm$ stands for part-per-million,  $f(t_n,v)$ is the integral of Equation \ref{eq:fluxmap_clar} over the disk, i.e. $j$ and $k$ indexes, and supposing a unitary total photospheric flux, while $f_{scale}$ accounts for data corrections and it has been chosen considering that the maximum model value, i.e. when spots are not present, should match the maximum possible value in the observed data. For this reason, we set $f_{scale}$ to $1+10^{-6} \cdot y_{max}$, where $y_{max}$ is the data maximum value of the observed tine series and the factor $10^{-6}$ arises from the data scaling which are expressed in ppm.
Considering all the assumptions we have done, the parameter vector ($v$) is made by the white noise jitter term ($\sigma_j$), latitudes \tbf{($\theta$)}, longitudes \tbf{($\phi$)} and radii \tbf{(R)} of the $m$ spot being considered. Prior values for the free parameters are presented in Table \ref{tab:priors}. Due to the defined assumptions on $i_{\star}$, the prior for the latitude has been restricted to the range \tbf{(0, 90)$\degree$}. This choice allows the presence of spots only in the northern stellar hemisphere and it avoids the degeneracy in the solution of two spots symmetric with respect the stellar equator.

Our strategy consists in running the retrieval repeatedly by increasing the number of spots (m) and by compare the Bayesian Evidence between two subsequent iterations. We take as best-fit solution the iteration which results in a difference in Bayesian evidence with respect the previous step of a least 5, considered as statistically strong evidence following \cite{1.3050814}. Subsequent fits will not be considered statistically relevant. To ensure a good sampling of the parameter space, we have selected 4000 live points in the \texttt{MultiNest} algorithm and to have a good compromise between model selection and parameter exploration, the sampling efficiency has been set to 0.5. Moreover, multimodal search has been set to True to allow efficient sampling in case of the presence of multimodal distributions while all other MultiNest parameters are left to their default values. To remove from posteriors initial and low probability values, a cleaning procedure have been applied to the results by selecting the samples with weight over the 0.75 quantile of the weight distribution. 

\begin{table}[]
    \caption{Prior list of the model parameters.}
    \label{tab:priors}
    \centering
    \begin{tabular}{c|c}
        \hline
        Parameter & Prior \\
        \hline
         $\sigma_j$ (ppm) & $LU(10^{-2},10^3)$ \\
         $\theta$ ($\degree$) & $U(0,90)$\\
         $\phi$ ($\degree$) & $U(-180,180)$\\
         R (R$_{\star}$) & $U(0,0.3)$\\
         \hline
    \end{tabular}
    
\end{table}

\subsection{Results}

As explained in previous subsections, the retrieval has been performed starting from considering only one spot and then repeated by increasing their number. Moreover, since we want not only to explain the photospheric variability but also to validate \texttt{PAStar}, we compare the surface configuration and the light curve from the model with the actual Sun spot configuration at similar times. To do this, we have selected the available images of the Sun, in the date of the selected VIRGO observation, from the Helioseismic and Magnetic Imager (HMI) \citep{2012SoPh..275..229S} on board of the Solar Dynamic Observatory (SDO) mission \citep{2012SoPh..275....3P} publicly available at the SDO/NASA database\footnote{https://sdo.gsfc.nasa.gov/data/aiahmi/}. Within this database, the HMI continum images have been downloaded, which scans the FeI absorption line at 6173 $\AA$ with a FHWM of 75 m$\AA$. Although the images are in a slightly different band with respect the temporal series to be fitted, we do not expect relevant differences in spots size and position and we made use of them only for comparison purposes.

\begin{table}[]
    \caption{Bayesian log evidence (logZ) and errors, as a function of the number of spots.}
    \label{tab:comp_bayes}
    \centering
    \begin{tabular}{c|c}
     \hline
      Spots  & LogZ  \\
      \hline
       1  &    -613.5  $\pm$  0.1 \\
       2  &    -587.2  $\pm$  0.1 \\
       3  &    -605.6  $\pm$  0.1 \\
       
       \hline
    \end{tabular}
    
\end{table}

\begin{figure*}
    \centering
    \includegraphics[scale=0.38]{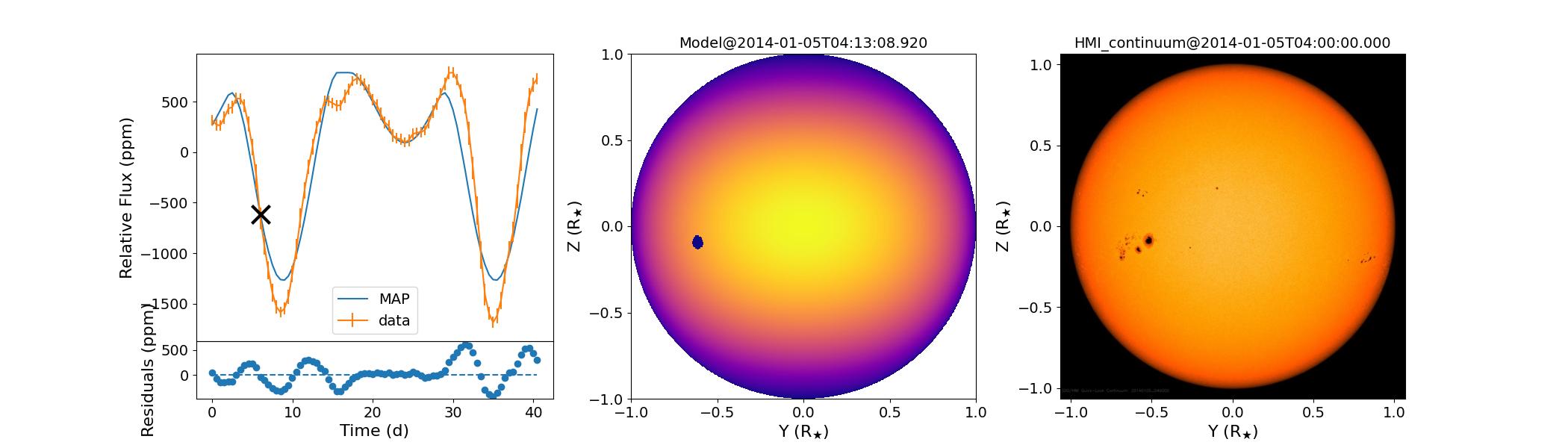}
    \includegraphics[scale=0.38]{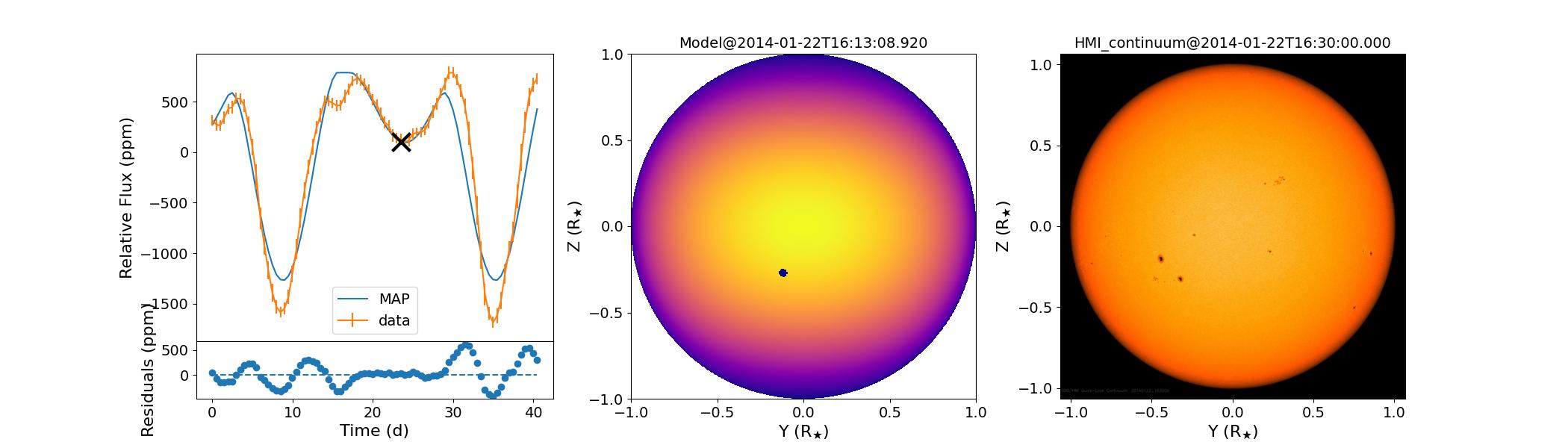}
    \caption{Comparison between the best-fit (MAP) model of the two spot configuration vs Sun data. (Left panels) Synthetic photometric light curve is presented as blue line together with the binned solar data as a orange line with its errors. A black cross marks the time at which the model (middle) and actual Sun (right) spot configurations have been shown. Times have been shifted with respect the first of the series. Spots position has been reflected with respect the equator for visualisation purposes \tbf{and stellar flux has been coloured-coded as for Fig.\ref{fig:comb-geom}}.}
    \label{fig:val_sun}
\end{figure*}

\begin{figure*}
    \centering
    \includegraphics[scale=0.35]{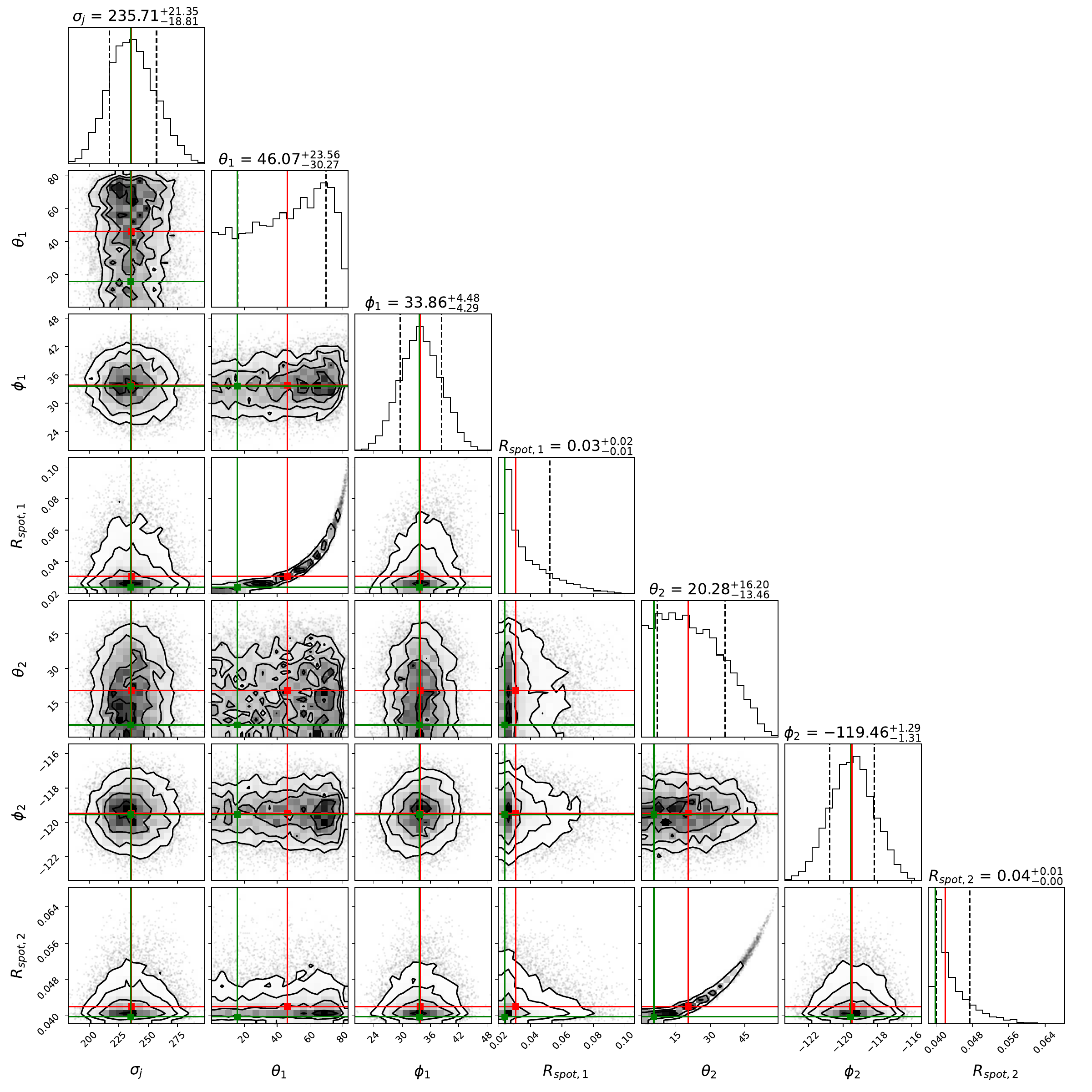}
    \caption{Posterior distribution of the two spot model. MAP solution is marked as as green line while the median as a red line. Dashed black lines in diagonal histograms mark, respectively, the 16$^{th}$ and 84$^{th}$ quantiles.}
    \label{fig:val_sun_corn}
\end{figure*}

The Bayesian log evidences (LogZ) for the different models we have tested is presented in Table \ref{tab:comp_bayes}. As a result, we have selected the configuration with 2 spots to better describe the data since represent the best evidence among the explored cases. The last tested scenario, i.e. 3 spots model, has a LogZ lower than the 2 spot model, therefore, we stop to increase the complexity of the model here, i.e. the number of spots. In Fig.\ref{fig:val_sun} we present the best-fit model considering the Maximum A-Posteriori probability (MAP) values for parameters together with the model spot configuration and the Sun spot configuration at similar times of HMI continuum images. The corresponding posterior distribution is presented in Fig.\ref{fig:val_sun_corn}. 

The solution found by the fit consists of two well defined spots which take turns in the visible disk. Spots longitude is retrieved with high precision even for the smaller spot. However, latitude are less constrained and the posterior indicates a band where spots can be found rather than a precise and well located position, as for the longitudes. This is not surprising because, while the longitude defines how the spot configuration modulates the photometric variation due to the rotation, the latitude, due to the limb darkening effect, controls the amount of the variation imposed by the spot presence. This effect should correlate with the size of the spot, therefore, we would expect a correlation between the latitude and radius parameters and this is exactly what the posterior shows. On other parameter which could correlate with radius and latitudes would be the spot temperature, however, here is set to zero. 

We report a difference in the shape of the synthetic profile with respect the observed one in the two main deeps. This effect could arise from the simplified assumption of circular spots which in this case can not reproduce the highly irregular shape of the observed spots and it could be particularly crucial for the Sun in which spots are small and irregular, at best of the size shown during the selected observation, and less evident for active stars in which their size increases. This discrepancy is overcame by high values of the white noise jitter which has a MAP comparable with the standard deviation of residuals ($\sigma_{res}=\sim240$ ppm). Another possibility to explain this discrepancy is the neglect of the spot flux together with the limb darkening could contribute to characterize the variation of the photometric profile. However, to find the temperature of sunspots is beyond the scope of this work in which the Sun is treated as a star in this validation process.

\subsection{Applicability of the model}
\begin{figure*}
    \centering
    \includegraphics[scale=0.38]{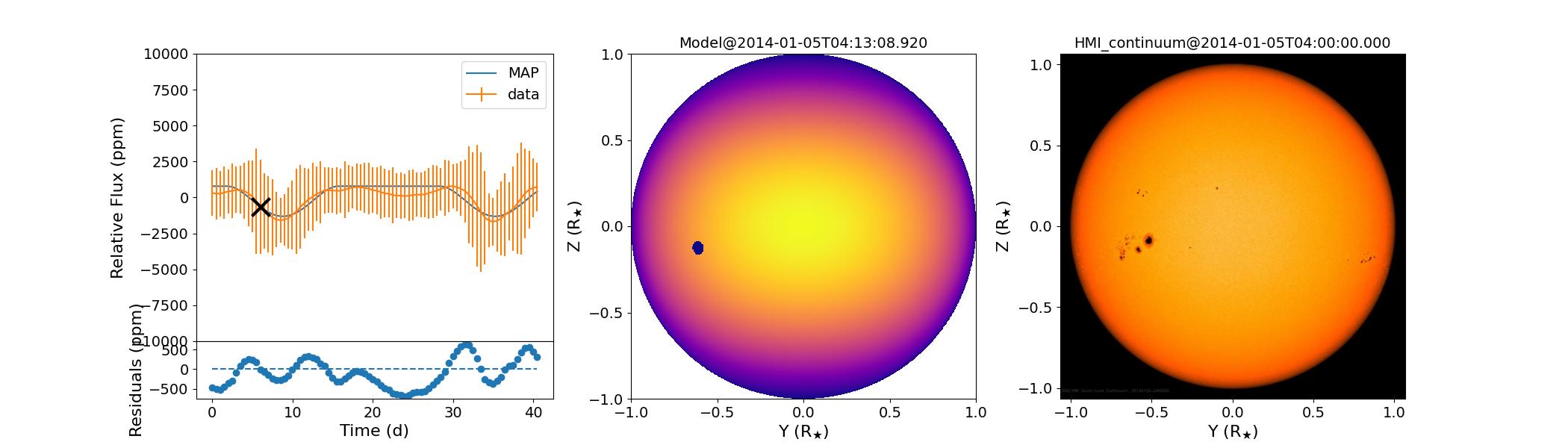}
    \caption{Same as for top panel of Fig.\ref{fig:val_sun} but for the low SNR case when one spot is considered.}
    \label{fig:val_sun_lwsnr}
\end{figure*}

\begin{figure*}
    \centering
    \includegraphics[scale=0.35]{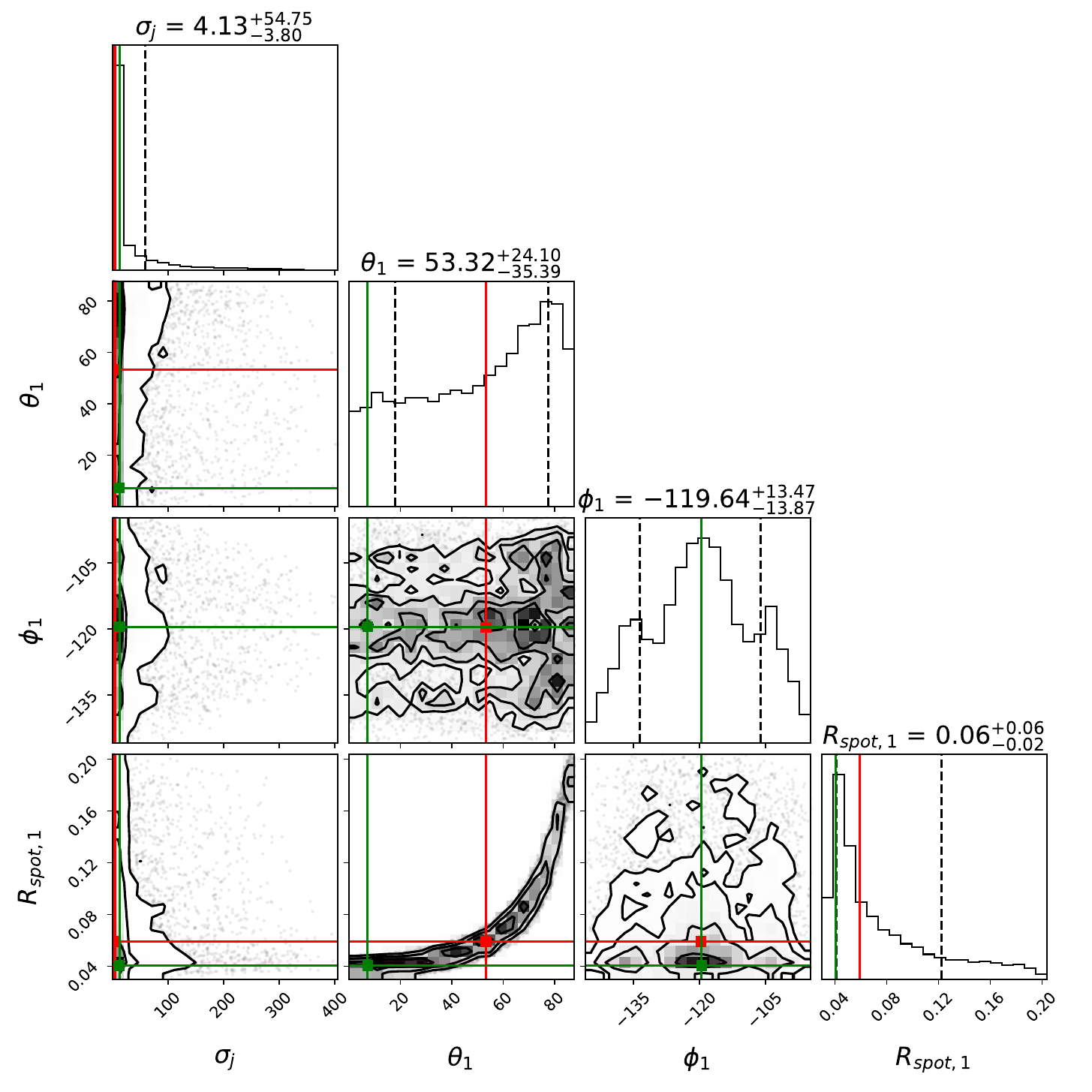}
    \caption{Same as for Fig.\ref{fig:val_sun_corn} but for the low SNR case.}
    \label{fig:val_sun_corn_lwsnr}
\end{figure*}

\begin{figure*}
    \centering
    \includegraphics[scale=0.38]{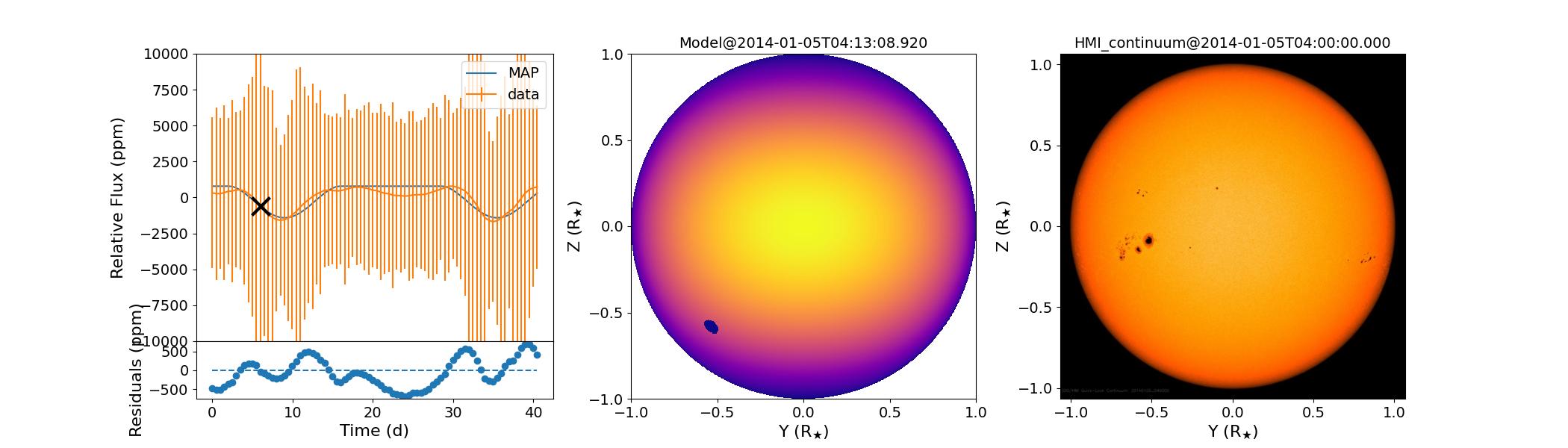}
    \caption{Same as for top panel of Fig.\ref{fig:val_sun_lwsnr} but considering a lower SNR, i.e. 100 times the data errors.}
    \label{fig:val_sun_lwsnr100}
\end{figure*}

In the previous subsection, the Sun has being considered as a star from a methodology point of view. We have derived the surface inhomogeneities reconstruction as we would apply the method to a far star in which we do not know spot temperature or star inclination. We have to note that \tbf{in solar data systematic errors dominate on measurements ones due to a very high signal-to-noise ratio (SNR), evaluated as $<y/\sigma>$, where $y$ is the stellar signal (flux) and $\sigma$ is its error}. Here, we have tested the model with all the assumptions above but in much lower SNR scenario, which could mimic that one of stellar observations. The question we have addressed is, therefore, what is the ability of our method in retrieving the spot configuration when the SNR decreases to levels of the typical photometric observations from space \citep[ex. Transiting Exoplanet Survey Satellite,][]{2015JATIS...1a4003R}. 
To artificially decrease the SNR we have multiplied data errors by a factor of 30 and repeated the iterative procedure explained in Section \ref{subsec:retrf}. In Fig. \ref{fig:val_sun_lwsnr} we present the result of this process.
In this exercise, the solution considering one spot is the most statistically significant, only the bigger spot is retrieved with a broader constrain in the latitude parameter, with respect the previous case\tbf{, as can be seen from the posterior distribution presented in Fig. \ref{fig:val_sun_corn_lwsnr}}. We have also tested other SNR scenario, however, for factors larger than 30 the MAP parameters retrieved do not recover correctly the actual position of the spot on the solar disk. \tbf{Therefore, this configuration is about the minimum activity level that \texttt{PAStar} can analyse due the combination of small filling factor of sunspots and the imposed SNR. For the sake of completeness, in Fig.~\ref{fig:val_sun_lwsnr100} the retrieved solar configuration increasing the error by a factor 100 has been presented. The difference between the actual latitude of the spot with respect the retrieved one increases with the increase of data errors, and the posterior distributions (not shown here) appear broader than the previous case, particularly in the longitude parameter.}

\section{Application on active stars - a faculae dominated scenario}\label{sec:active star}

In this section we test the model and the retrieval framework against an activity scenario dominated by faculae. The synthetic stellar configuration has been retrieved following the methodology described in Section~\ref{sec:valid} that considers an iterative procedure to determine the number of appropriate inhomogeneities on the stellar surface and, thereafter, the surface configuration. In this approach, the same retrieval process is derived for an increasing number of inhomogeneities which alts when bayesian evidence of the best-fit solution does not increment more than 5 with respect the solution of the previous step.
As photometric light curve generated by \texttt{SOAP}, the Equatorial-view case (i$_{\star}$=90$\degree$) of the synthetic light curves presented in Section~\ref{sec:soap_comp} has been selected, which represents the case of an active star. This configuration employs 4 faculae distributed across the northern stellar hemisphere and whose characteristics are listed in Table \ref{tab:synth-faculae}. We derive this validation in two different scenarios:\\
\begin{enumerate}
  \item we assume to know that the star is faculae dominated and therefore we suppress the presence of spots on the surface (Scenario 1);
  \item we do not have enough information on the star to formulate a hypothesis and consequently we search for a general spot and faculae configuration (Scenario 2).
\end{enumerate}

\begin{figure*}
    \centering
    \includegraphics[scale=0.38]{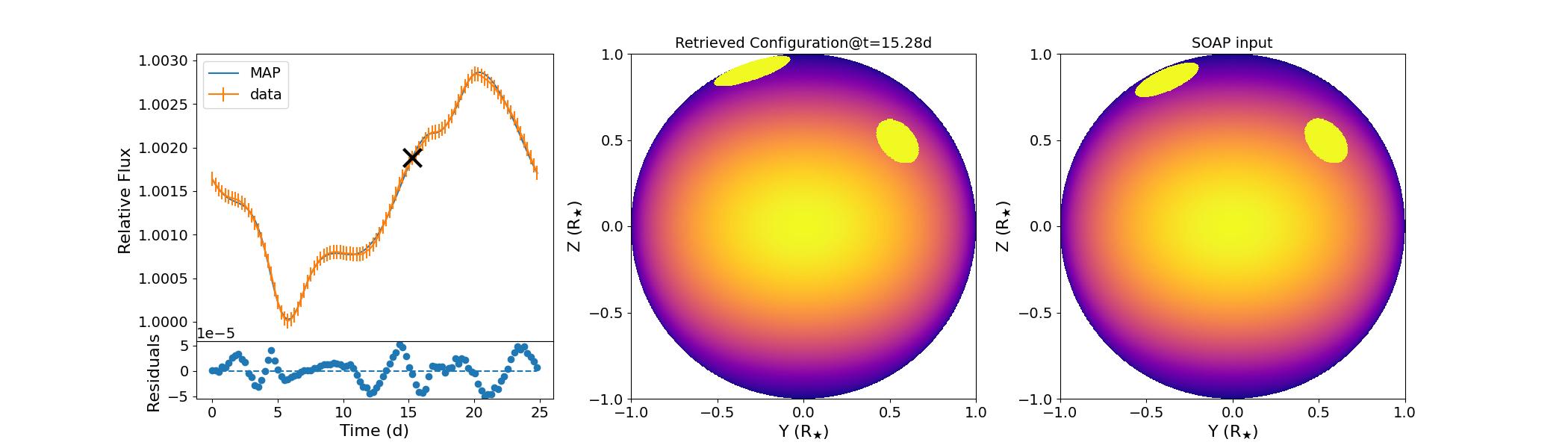}
    \includegraphics[scale=0.38]{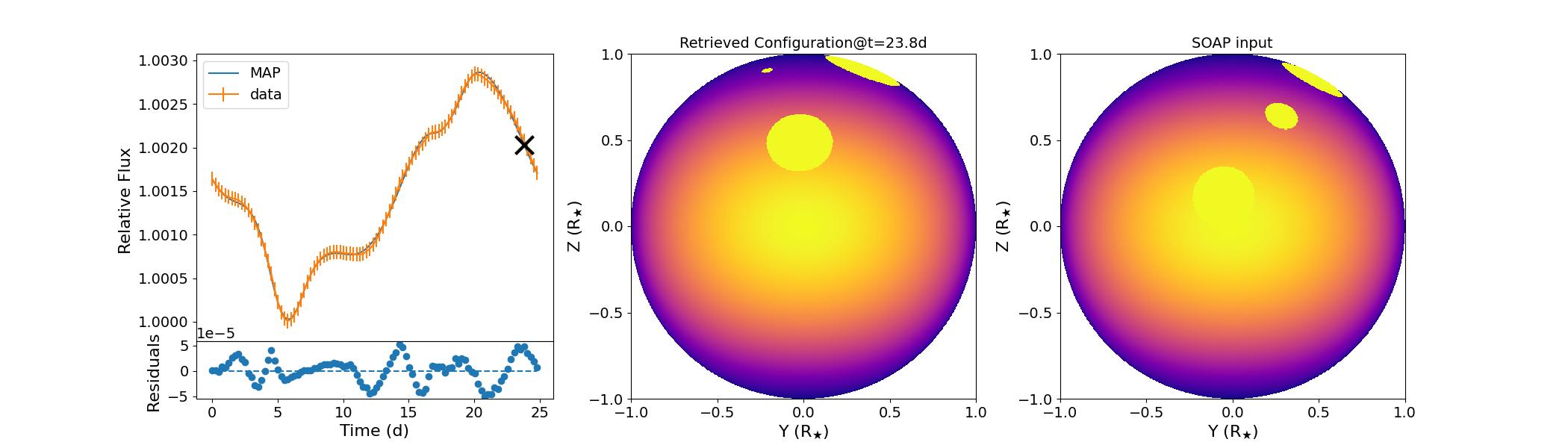}
    \caption{Comparison between the best-fit (MAP) model of the 4 faculae configuration vs input \texttt{SOAP} configuration. (Left panels) Best-fit photometric light curve is presented as blue line together with the SOAP synthetic light curve as a orange line with its errors, obtained adding in quadrature data errors and jitter. A black cross marks the time at which the model (middle) and input \texttt{SOAP} (right) faculae configurations have been shown. \tbf{Stellar flux has been coloured-coded as for Fig.\ref{fig:comb-geom} with the exception of the faculae flux that has been modified to 10$^3$ in order to enhance the visibility.}}
    \label{fig:synth-faco-retr}
\end{figure*}

\begin{figure*}
    \centering
    \includegraphics[scale=0.38]{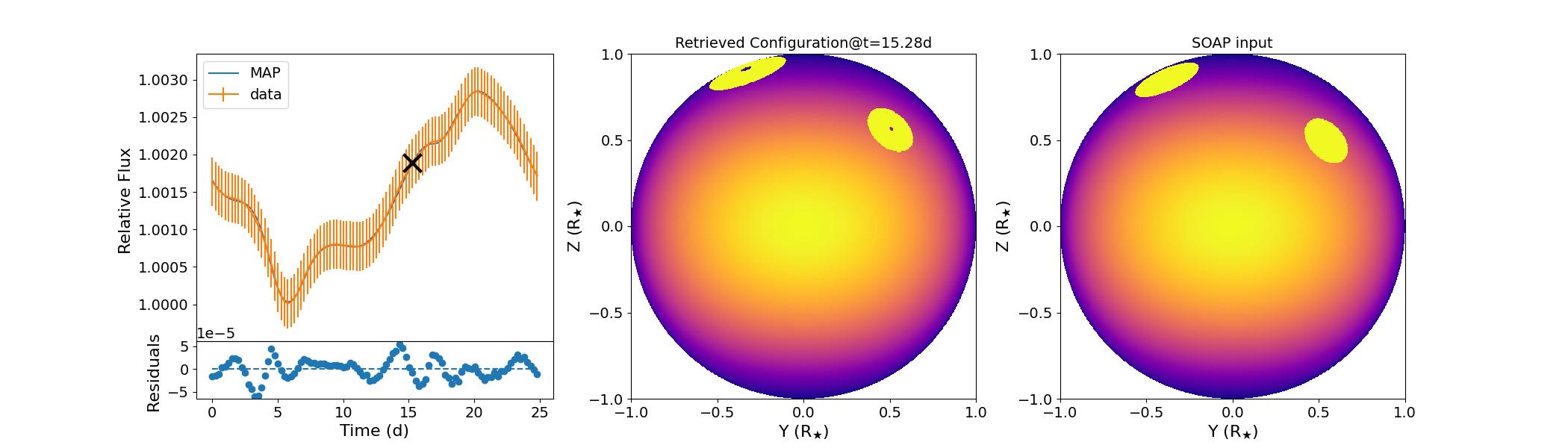}
    \includegraphics[scale=0.38]{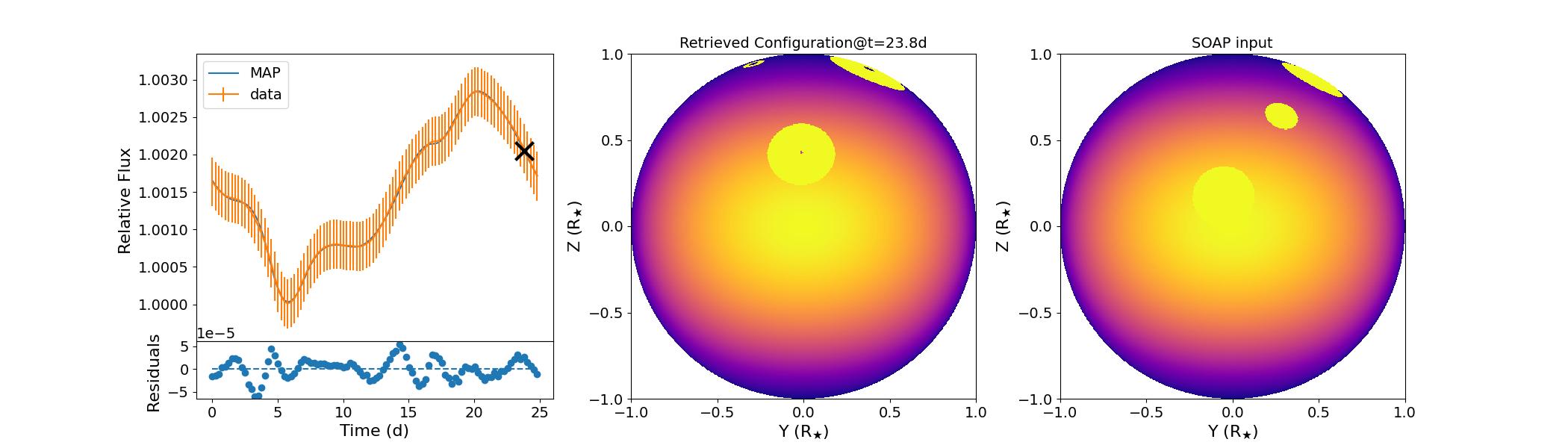}
    \caption{Same as for Fig.\ref{fig:synth-faco-retr} but in the case of the Scenario 2.}
    \label{fig:synth-spot+faco-retr}
\end{figure*}

\begin{figure*}
    \centering
    \includegraphics[scale=0.25]{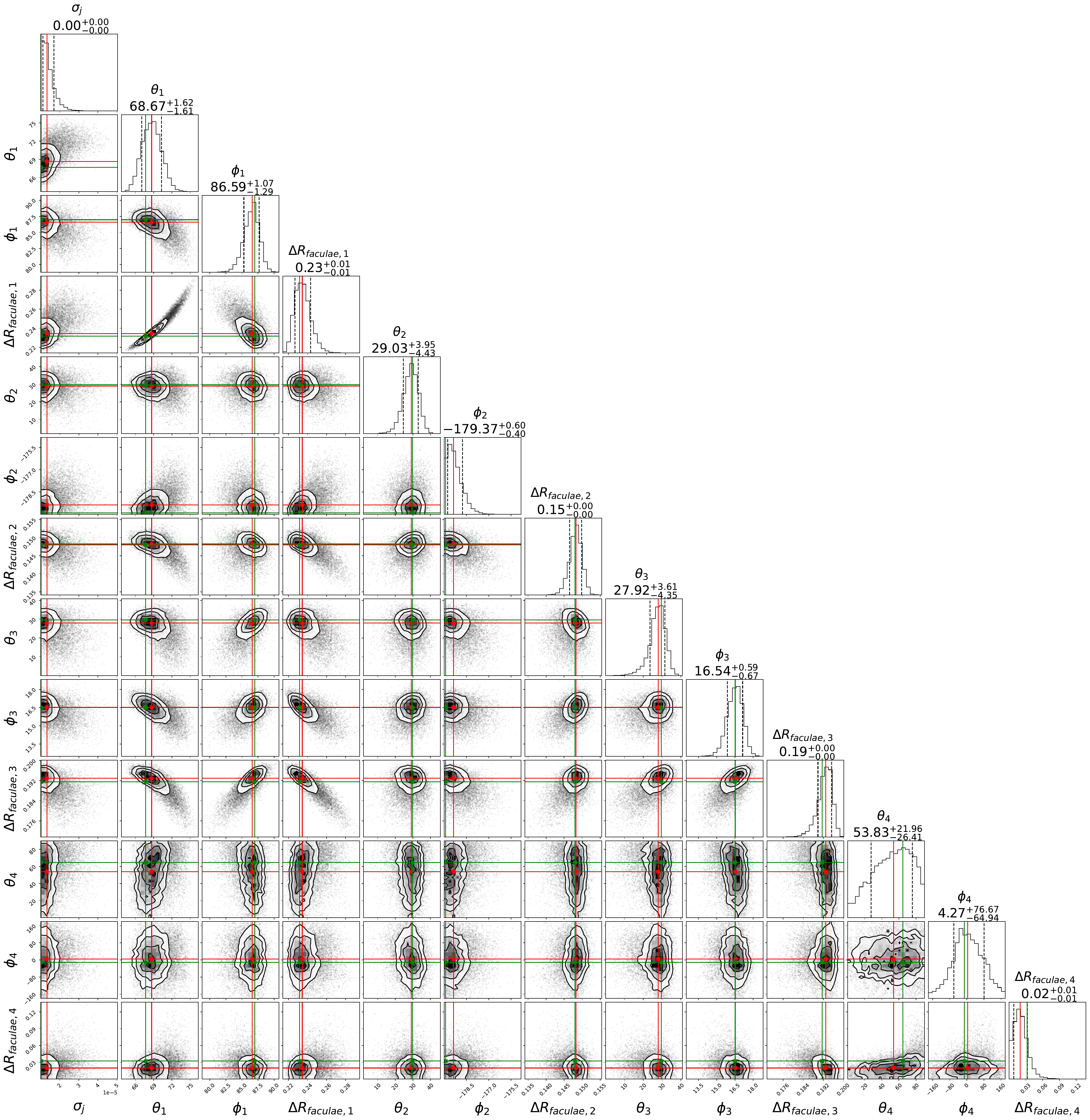}
    \caption{Same as for Fig.~\ref{fig:val_sun_corn} but in the case of the faculae-dominated Sun of the validation process presented in Fig.~\ref{fig:synth-faco-retr}.}
    \label{fig:synth-faco-retr-corn}
\end{figure*}

In both scenarios, we have retrieved the surface configuration by varying, arbitrarily, data error ($\sigma$) between 10\% and 40\% of the photometric standard deviation, in order to study limits and degenerations of the retrieval process in two SNR cases. Bayesian evidences of these retrievals are presented in Table~\ref{tab:synth-faculae-retrieval}. The best-fit configuration in the two scenarios are presented, respectively, in Fig.~\ref{fig:synth-faco-retr} and Fig.~\ref{fig:synth-spot+faco-retr} while posterior distributions in the case of the best-fit solution for Scenario 1 are presented in Fig.~\ref{fig:synth-faco-retr-corn}.

\begin{table}[]
    \caption{Bayesian log evidences (logZ) and errors, as a function of the number of inhomogeneities added in the retrieval, in the two tested scenarios.}
    \label{tab:synth-faculae-retrieval}
    \centering
    \begin{tabular}{c|c|c|c|c}
     \hline
     & \multicolumn{2}{|c|}{Scenario 1}  & \multicolumn{2}{|c}{Scenario 2} \\
     \hline
      Spots  & $\sigma_{10\%}$ & $\sigma_{40\%}$& $\sigma_{10\%}$ & $\sigma_{40\%}$ \\
      \hline
       1  & 701.0 $\pm$ 0.2 & 680.8 $\pm$ 0.2 & 699.6 $\pm$ 0.2 & 678.7 $\pm$ 0.2\\
       2  & 720.0 $\pm$ 0.2 & 681.9 $\pm$ 0.2 & 715.7 $\pm$ 0.2 & 676.9 $\pm$ 0.2\\
       3  & 785.0 $\pm$ 0.3 & - & 744.5 $\pm$ 0.3 & - \\
       4  & 797.2 $\pm$ 0.2 & - & 790.5 $\pm$ 0.2 & - \\
       5  & 799.0 $\pm$ 0.2 & - & 774.3 $\pm$ 0.3 & - \\
      \hline
    \end{tabular}  
\end{table}

As a result of the bayesian evidences comparisons, the retrieval process favors the solution of Scenario 1 that matches the number of input faculae. Moreover, all the solution of Scenario 1 are statistically more significant than their counterparts in Scenario 2 by giving systematically higher bayesian evidences. In the case of lower SNRs, in both scenarios the best-fit solutions are characterized by only one inhomogeneity with no match with respect the input solution. 
The obtained best-fit solution in the two scenarios (Fig.\ref{fig:synth-faco-retr} and Fig.\ref{fig:synth-spot+faco-retr}) are similar in terms of surface configuration but differ in the presence of very small spots in Scenario 2 which never exceed 0.04 R$_\star$ and act as a source of noise by increasing the data jitter and, consequently, resulting in less constrained posterior distribution (not shown here). 
In general, the obtained surface configuration resembles the input one but some discrepancy is present, as evident from lower panels of Fig. \ref{fig:synth-faco-retr}-\ref{fig:synth-spot+faco-retr}. The smallest faculae ($\Delta$R$_{faculae}$ = 0.1 R$_{\star}$) of the input configuration is missing from the obtained solution in favor of a non-existent inhomogeneity at a high latitude ($\theta$ $\sim60\degree$) but with a smaller radius. The effect of the missing inhomogeneity has been overcame by the model in placing the input equatorial faculae at an higher latitude to average the position of two close inhomogeneities. This behavior is the same as the one shown in Section~\ref{sec:valid} where spots average the position of the actual spots on the Sun photosphere. Here, this effect apply to faculae which are separated by less than $\sim30\degree$ (center-to-center). When faculae separation is higher, the correct position is retrieved, within a conservative 5 $\sigma$ as obtained from the posterior distributions.

\begin{figure*}
    \centering
    \includegraphics[scale=0.38]{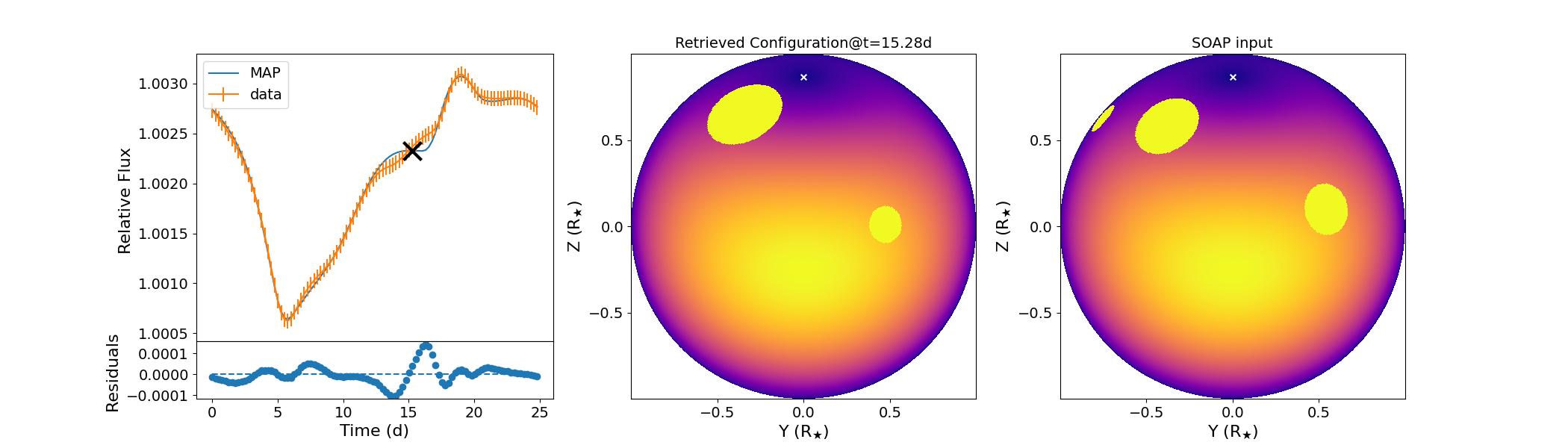}
    \includegraphics[scale=0.38]{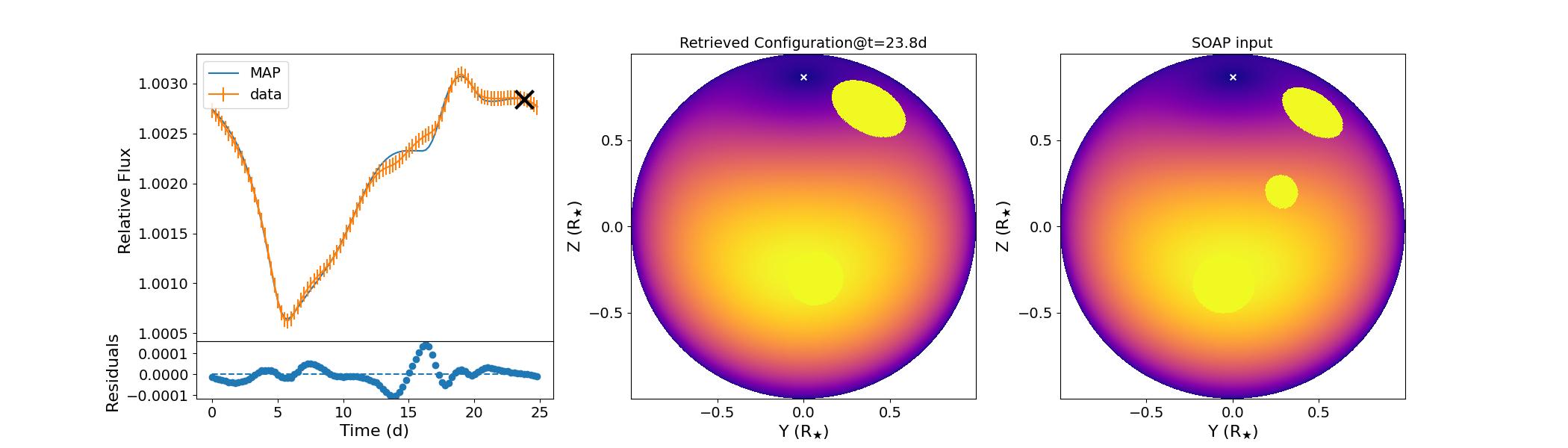}
    \caption{Same as for Fig. \ref{fig:synth-faco-retr} but considering i$_{\star}$ = 60$\degree$. A white cross marks the pole of the star. }
    \label{fig:synth-faco-retr60}
\end{figure*}

\begin{figure*}
    \centering
     \includegraphics[scale=0.25]{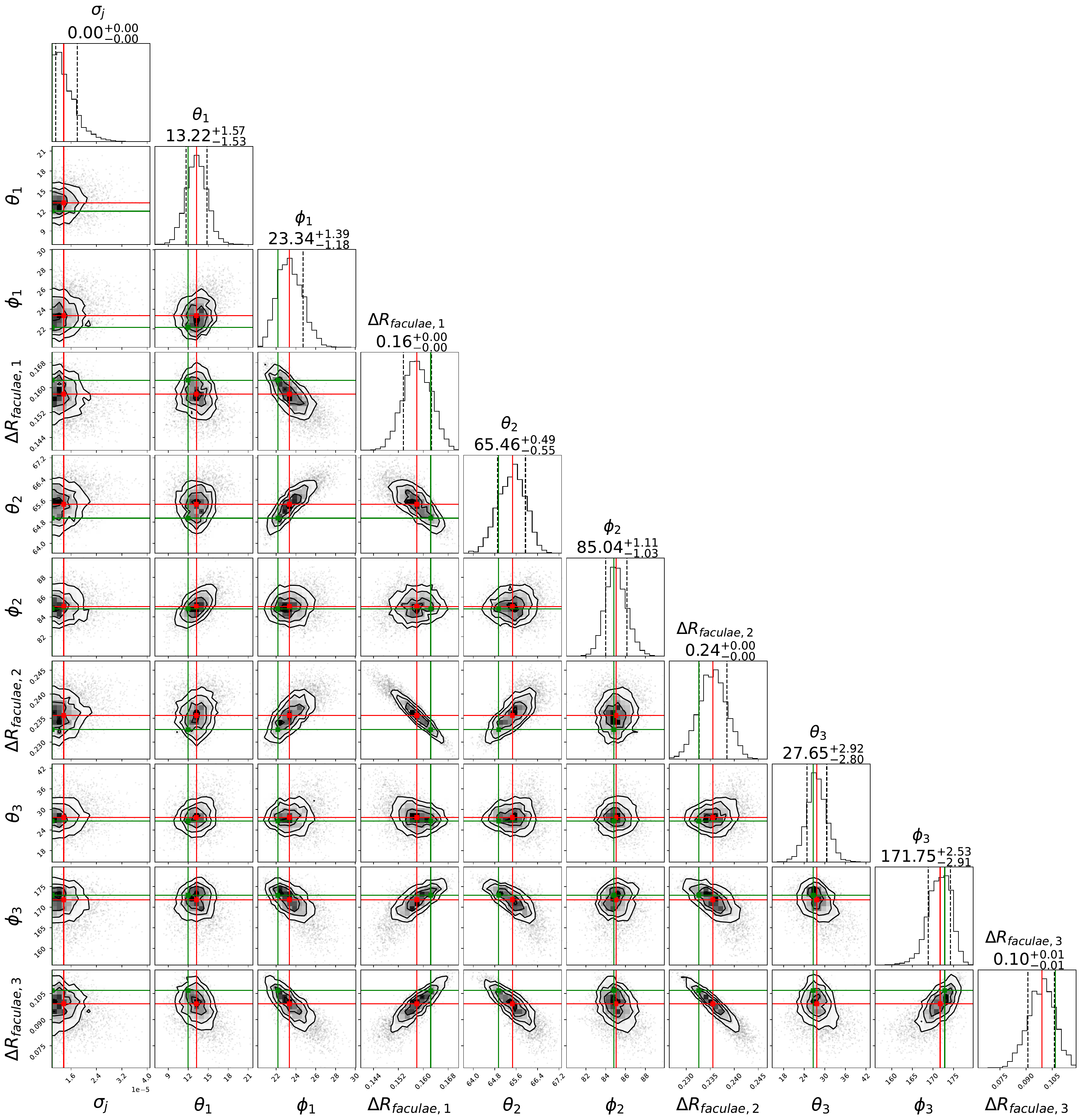}
    \caption{Same as for Fig.~\ref{fig:synth-faco-retr-corn} but considering i$_{\star}$ = 60$\degree$. Although multiple modes were present in the posterior distributions, only the one containing the MAP has been shown and used to evaluate uncertainties.}
    \label{fig:synth-faco-retr-corn60}
\end{figure*}

\tbf{We have extended this application to the case of i$_{\star}$=60$\degree$, but only for Scenario 1 and for the highest SNR, i.e. considering data errors equal to 10\% of the standard deviation of data. The result of the retrieval and the posterior distribution are presented, respectively, in Fig. \ref{fig:synth-faco-retr60} and Fig. \ref{fig:synth-faco-retr-corn60}. Differently from the previous case, although the 4 faculae best-fit solution has a greater bayesian evidence ($\log Z=806.5$) than the 3 faculae case ($\log Z=801.7$), following our selection criterion we choose the latter as the most significant one. As a consequence, the smallest faculae is missing from the reconstructed surface with the others that are correctly retrieved within a conservative 10 $\sigma$.}

\section{Discussion and Conclusions}\label{sec:disc}
In this work, we present a model to describe the stellar photosphere on particularly active stars. They are characterized by surface inhomogeneities, spots and faculae, whose define the observed variability. We model the stellar surface as superposition of components, photosphere, spots and faculae. They are spherical cap whose shape is being distorted by their projection on the stellar surface and they are described throughout independent stellar atmosphere.

The model describes the fundamental effects to be taken into account in order to describe the observed variability. These effects are the stellar inclination with respect to the line of sight, the limb darkening, independent for each component, and the broadening of spectral line shape due to the stellar rotation. \tbf{It differentiates from other models present in literature, primarily, for the geometrical construction and the flexibility in describing different kind of inhomogeneities in the same stellar surface, without any assumption on their radii.}

The model produces several products as photometric and spectroscopic light curves, also in presence of a transiting exoplanet, and for which the occulted flux by the planet is evaluated to allow further studies regarding the exoplanet atmosphere. This variety allows the model to be applied to several scientific cases thus, expanding its applicability.

To check for the consistency of our synthetic products, a comparison with an analogous model present in literature has been made. To this purpose, a synthetic faculae-dominated Sun has been generated with the \texttt{SOAP} code \citep{Dumusque_2014}, for 3 stellar inclinations. We report a good match between solutions generated by the two codes with residuals within 0.004\% with respect the \texttt{SOAP} solution which can be relate to the different geometrical construction. \tbf{We have shown that our model gives smooth and reliable light curves right at low resolution while \texttt{SOAP} code needs a much higher number of grid points.}

The model has been tested against solar data in order to validate the geometrical construction and the evaluation of the photometric light curve. Although the Sun is not a particularly active star it is the only object in which this validation could be made uniquely. For this reason, it offers a great challenge for the validation because of the dimension of the spots which is generally small compared to other active stars.  

By combining the model to a retrieval framework we are able to constrain \tbf{latitude, longitude and radius of the} spots on the solar surface. We find that a model with at least two spots is able to explain most of the observed variability for the specific temporal window chosen, obtaining a good match between the modelled versus the actual spots location and size. Due to the simplicity of the spot geometry, modelled spots represent an 'effective' spot of the conglomerates observed. This effect arises differences in the modelled light curve with respect to the observed and it is enhanced by the small filling factor of the observed spot. By testing the model in real SNR scenario, we demonstrate the ability of the model to retrieve spot information when spots are as small as $0.05$ $R_{spot}/R_{star}$. However, we stress that targets of applicability of the model are rather active stars in which spots size and their filling factors are much greater than the solar case and for those stars the effect of non circular spots would be less evident.

The validation process has been also extended to active faculae dominated stars by using synthetic\tbf{s} photometric light curve\tbf{s} generated by the \texttt{SOAP} code. The faculae configuration has been retrieved in two scenarios in terms of modelled inhomogeneities, i.e. faculae-dominated and spot+faculae scenarios, and in two SNRs cases, i.e. 10\% and 40\% of the photometric variability\tbf{, in the equatorial view case}. As a result, the retrieved configuration matches the input one, provided that the SNR is sufficiently high. However, although the input configuration is globally retrieved in the case of the facula-dominated scenario, discrepancy in faculae properties, between input and retrieved configuration, could be obtained when the input faculae are closer than 30$\degree$. This behaviour is also observed in the solar case, but in a smaller scale, where the retrieved spot is an average of the observed spot configuration. \tbf{We have extended this analysis to the case of $i_{\star}=60\degree$ in the faculae-dominated scenario and for the highest SNR. Also in this case, we are able to retrieve 3 faculae out of the 4 of the input configuration with only the smallest missing.}

The model does not include differential rotation which may be relevant depending on the stellar type being studied. In solar-type stars, equatorial rotation could differ significantly from the polar one. In the Sun this effect results in a difference of $\sim 10$ days \citep[ex.][]{Thompetal2003} between rotational periods.
As in the case of the Sun validation, when the effect of the differential rotation is neglected, but present, a different rotational period should be employed to describe correctly the photometric variability which reflects the rotational period of the active latitude band where inhomogeneities lie. In this validation process, it has been evaluated from data as the difference between minima.  
However, our primarily targets of interest are active stars in which the fast rotation limits the effect of differential rotation \citep{refId0} that although present could be subsequently lowered by limiting the analysis to short temporal windows, shorter than the equatorial rotational period.

Another physical assumption that could affect our results is the imposition of the spot and/or faculae temperature to values which could not be correct and, consequently, leading to an imprecise determination of latitude and radius of inhomogeneities, due to the strong degenerations expected.
This degeneracy could be break by a simultaneous multiband analysis that takes advantage of the different flux contrast ratio of inhomogeneities, with respect photosphere, in different spectral bands \citep{2012A&A...539A.140B}.

In conclusion, the model we have developed offers a flexible and valuable tool to describe the activity of stars when it is dominated by spots and faculae. It has a wide range of applicability due to the great variety of products. \tbf{The high flexibility in the inhomogeneities description allows our model to describe a wider range of activity cases with respect the models present in literature.}
The description of stellar activity is a fundamental step in several astrophysical contexts and it is covered by the method we have presented.
In term of future developments, the flexibility of the model allows us to add effects to the physics description of the stellar surface, as for the differential rotation of the star, spots and faculae temperature, with particularly ease, thus, enhancing the ability of the model to describe more scientific cases.

\begin{acknowledgements}
\tbf{The authors thank the referee for the useful comments and suggestions that improved deeply our manuscript.}
The authors acknowledge the support of the ASI-INAF agreement 2021-5-HH.0. J. M. acknowledges support from the Italian Ministero dell'Università e della Ricerca and from the European Union - Next Generation EU through project PRIN MUR 2022PM4JLH ``Know your little neighbours: characterizing low-mass stars and planets in the Solar neighbourhood''. A.P. acknowledge support from the INAF Minigrant of the RSN-2 nr. 16 "SpAcES: Spotting the Activity of Exoplanet hosting Stars" according to the INAF Fundamental Astrophysics funding scheme.
A. P. and G. M. acknowledge support from the European Union - Next Generation EU through the grant n. 2022J7ZFRA - Exo-planetary Cloudy Atmospheres and Stellar High energy (Exo-CASH) funded by MUR - PRIN 2022.\\ 

\end{acknowledgements}
\bibliographystyle{aa}

\bibliography{bib}

\end{document}